\documentclass[aps,epsf,preprint]{revtex4}

\def\1{{\bf 1}}
\def\[{\left[}
\def\]{\right]}
\def\be{\begin{eqnarray}}
\def\ee{\end{eqnarray}}
\def\nn{\nonumber}
\def\({\left(}
\def\){\right)}
\def\bk#1{\langle#1\rangle}
\def\eq#1{(\ref{#1})}

\def\a{\alpha}
\def\r{\rho}
\def\o{\omega}

\def\f{\phi}
\def\q{\psi}
\def\G{{\cal G}}
\def\C{{\cal C}}
\def\l{\lambda}

\def\x{\times}
\def\ket#1{|#1\rangle}
\def\bra#1{\langle #1|}
\def\vss{\vskip.1cm}
\def\p{\partial}
\def\d{\delta}

\def\labels#1{\label{#1}}
\def\edc{\end{document}}
\def\P{{\cal P}}

\def\bn{\begin{enumerate}}
\def\i{\item}
\def\en{\end{enumerate}}
\def\b{\beta}
\def\g{\gamma}
\def\ol{\overline}
\def\rd{\sqrt{2}}
\def\rt{\sqrt{3}}
\def\rc{\sqrt{5}}
\def\rs{\sqrt{6}}
\def\diag{{\rm diag}}
\def\th{\theta}
\def\ba{\begin{array}}
\def\ea{\end{array}}
\def\bc{\begin{center}}
\def\ec{\end{center}}

\begin{document}

\title{Group Theory and Dynamics of Neutrino Mixing}
\author{C.S. Lam}
\address{Department of Physics, McGill University\\
 Montreal, Q.C., Canada H3A 2T8\\
and\\
Department of Physics and Astronomy, University of British Columbia,  Vancouver, BC, Canada V6T 1Z1 \\
Email: Lam@physics.mcgill.ca}

\begin{abstract}
There is a direct group-theoretical connection between neutrino mixing and horizontal symmetry that can be established without any 
dynamical input. Such a connection is
reviewed and expanded in this article. For certain symmetry groups $\G$ including $A_4$ and $S_4$, it is shown that
a generic $U(1)\x\G$
Higgs potential of a valon  yields exactly the alignments dictated by the group-theoretic approach, 
but  energy can now be used
to discriminate different alignments. This mechanism possibly explains
why starting from an $A_4$ group, the tribimaximal mixing matrix with an enhanced $S_4$ symmetry is more
preferable  than the one without it. 
\end{abstract}
\narrowtext
\maketitle

\section{Introduction}
After Chadwick discovered the neutron in 1932, Heisenberg came up with the  `isotopic spin' to distinguish it from
the proton.
This new quantum number corresponds to a global symmetry $SU(2)$ which explains the similarity 
of neutrons and protons in strong interactions. Now that we have three generations of quarks
and leptons, should there not be a new `family quantum number' to distinguish them, and 
a corresponding `horizontal (or family) symmetry' to explain the similarity
of their interactions? 

Unlike neutrons and protons which have nearly the same mass, the masses of fermions of different
generations are vastly different. Moreover, they mix. Thus if horizontal symmetry is present,  it has to be badly broken, a
fact which makes the symmetry difficult to recognize unless enough remnants survive from the breaking to tell the story. We
suggest that such remnants are indeed present and they are hidden in the mixing matrix.
 
Neutrino mixing is successfully described by the tribimaximal mixing (TBM) matrix \cite{HPS}. This mixing is most
frequently explained by
models with an $A_4$ symmetry, but
models based on $S_4$  as well as many other groups $\G$ \cite{REVIEW} can also do the job. 
With so many successful models, it is hard to know which of them is the correct
horizontal symmetry, and indeed, whether horizontal symmetry really exists or not. 

To shed some light on this question from a  different angle, we have previously developed a purely group-theoretical 
method to connect neutrino mixing with horizontal symmetry \cite{LAM}. This method differs from the conventional one in that only symmetry considerations are involved. The presence of Higgs fields is never assumed, nor the details of a model Lagrangian.
In this approach, the broken symmetry (which we call 
{\it residual symmetry}) and the  minimal
unbroken horizontal symmetry
of the {\it left-handed leptons} are derived from the mixing matrix $U$. 
In particular, if neutrino mixing is given by the TBM, then the smallest unbroken horizontal symmetry
is $S_4$, not $A_4$. Models based on $A_4$ can explain TBM because in those models
there is an accidental symmetry which elevates the $A_4$ symmetry to an $S_4$ symmetry.
These results will be reviewed in Sec.~II.

The same formalism can in principle be applied to fermion mixing as well \cite{LAM}, but unless we make suitable approximations to 
the CKM mixing, all that we get is the group $SU(3)$, nothing smaller. For that reason we will not discuss quark mixing any further in
this paper. This difference, however, brings up a very important question, as to whether there is a common origin between neutrino mixing
and quark mixing that is smaller than $SU(3)$. This question will be taken up in a forthcoming paper \cite{LAMNEW}.

The group-theoretical procedure  can be reversed  to obtain mixing matrices $U$
from a symmetry group $\G$. The detail will be discussed in Sec.~III, but let us outline now how that differs from the usual
dynamical method. In the dynamical approach, $U$ is to some extent determined by the vacuum alignments, 
and  the  alignments are in principle obtained by minimizing a Higgs potential $V$. 
There are however frequently too many $\G$-invariant terms available to construct $V$, so 
 that just about any vacuum alignment can be obtained by a suitable choice of $V$. 
What one does then is to use the experimental results to decide what $U$ one wants, and what alignments that can give rise to such an $U$, then proceeds to
design a $V$ that does the job. To keep it robust, this is often accomplished by imposing another symmetry in additional to $\G$
to throw away the unwanted terms   in $V$. 

In the group-theoretical approach, $U$ is determined by the residual symmetries $F$ (for the charged-lepton sector)
and $G_i\  (i=1,2,3)$ (for the neutrino sector) which are members of $\G$. No peeking of data is necessary
 so one is not prejudiced
by the experimental outcome. There is no dynamics to tune, and the existence of Higgs is not assumed. For a given $\G$, there is usually more than one possible
choice of residual symmetries and thus more than one $U$, but within equivalence only very few survives. By equivalence I mean the following.
 Since Majorana phases are not measurable in neutrino oscillation experiments, and since
there is no way for group theory to distinguish which flavor neutrino is which, and which neutrino mass eigenstates are 
to be labeled 1, 2, or 3, we will consider two mixing matrices equivalent if they differ only by row and column phases, and by row and column permutations.
With a convenient choice of permutation, the mixing can be specified by three mixing angles and one CP phase, contained in the triplet 
$\bk{\sin\th_{12},\sin\th_{23},\sin\th_{13}e^{-i\d}}$. Two $U$'s with the same triplet parameters are considered to be equivalent.

Normally we need a $F$ and three mutually commuting $G_i$ to specify a $U$. A mixing matrix determined this way will be called a
{\it full-mixing} matrix. These are the mixing matrices obtained from $\G$ by symmetry alone.
In order to allow TBM to occur under $A_4$, we will also discuss {\it partial-mixing} matrices, which are obtained from
a full-mixing matrix by replacing two of its columns with arbitrary parameters, subject however to the constraint of unitarity. 
Equivalently, they are mixing matrices determined by $F$ and one of the three $G_i$'s, plus additional parameters needed to specify $U$.
For each full-mixing there
is always associated three types of partial-mixings, which will normally not be mentioned separately. 

However, the mixing matrix obtained from a partial-mixing  of $\G$ with the other two columns filled  has its own symmetry group
$\ol \G$ which contains $\G$ as a subgroup. Unless the entries in the other two columns are carefully chosen, most likely $\ol\G$ would end
up to be an infinite group, perhaps as large as $SU(3)$. Thus, the entries of the other two columns are severely restricted if we stipulate 
$\ol\G$ to be a finite group. In particular, if the entries are those given by the full-mixing matrix, then $\ol\G=\G$. In the case
of $A_4$, the smallest finite group containing it is $S_4$, and there is one partial-mixing mode of $A_4$ whose $\ol\G$ is
$S_4$. That is how TBM is produced from $A_4$.

In the case of $S_4$, there are {\it two} full-mixing patterns. One is the familiar TBM specified by the physical parameters $\bk{{1\over\rt},{1\over\rd},0}$, and the 
other is one specified by $\bk{{1\over\rd},{1\over\rd},0}$, which is seldom mentioned in the literature. The full-mixing
of $A_4$ is the Cabibbo-Wolfenstein mixing matrix, specified by $\bk{{1\over\rd},{1\over\rd},{e^{-\pi i/2}\over\rt}}$,
though TBM is allowed in a partial-mixing mode.
There are two possible full-mixings in $A_5$, but we will work out only  one whose physical parameters are $\bk{{1\over\rt \varphi},
{1\over\rd},0}$, where $\varphi=(1+\rc)/2=1.618$ is the golden ratio. Once again, these results are obtained in a purely group-theoretical way,
without the help of any dynamics.

Secs.~IV, V,  VI are devoted to the connection between  group-theoretical  and  dynamical approaches. Defining an {\it invariant eigenvector} to be an
eigenvector with eigenvalue $+1$, the basic connection is 
that the vacuum alignment in every IR has to be the invariant eigenvector of the corresponding residual
symmetry operator  in that IR \cite{LAM}. That connection is reviewed in Sec.~IV.
In case such an eigenvector does not exist,
the  alignment would be forced to be zero and the coupling with such a valon cannot contribute. This in principle
 would reduce our ability to make models
because we need  at least
three independent Yukawa couplings in each sector to fit the three charged-lepton and the three neutrino masses.
Nevertheless, in the case of $S_4$ and $A_4$ which we discuss in some detail, there remain just enough parameters to fit all the data
in  type-II seesaw models.
For $A_4$ we also discuss the partial-mixing results, both because the phenomenologically interesting TBM belongs to
that category, and because the result of the discussion would be useful in Sec.~V.
 We see from these discussions that a large number of models can be constructed for each group, and indeed,
many of them have already appeared in the  literature. Similar discussions can be carried out
for other groups.

With the basic connection, one important question to ask  is whether  a generic Higgs potential 
can be constructed to yield just these vacuum alignments. As remarked before,
in order to avoid the appearance of arbitrary alignments, an additional symmetry would have to be imposed to throw away unwanted
 $\G$-invariant terms.
In the case of $\G=A_4$, that additional symmetry turns out to be $U(1)$ (an appropriate $Z_n$ would do as well) as we shall show in Sec.~V. The calculation involves writing down a generic $U(1)\x A_4$ potential in terms of CG coefficients in a convenient basis,  then
minimizing it. 

Can this result be generalized to other groups? 
On the one hand, 
this calculation requires intimate knowledge of the group $A_4$, such as the explicit CG coefficients and the details of
all the ($U(1)\x A_4$)-invariant terms, so it is difficult to see how to generalize it from $A_4$ to other groups $\G$. 
On the other hand,
the result of minimization can be expressed in general group-theoretic terms, as the invariant eigenvectors of members of $A_4$, and that
gives hope that one might find a way to bypass the details of $A_4$ to generalize the result to other groups $\G$. This is indeed
true as will be discussed in Sec.~VI. In the first part of that section, an alternative proof for the $A_4$ result is presented that
is quite independent of the specific $A_4$ details. This proof is then generalized to other groups $\G$ in the rest of that section.

Besides demonstrating the consistency between the group-theoretical and the dynamical approaches, 
the introduction of the Higgs potential also serves another purpose.
There is no way for group theory to choose between different residual symmetries, but Higgs potentials might, as different
vacuum alignments might give rise to different Higgs energy. In that case the one with the lowest Higgs energy wins out.
 
This mechanism can for example be used to solve a puzzle.
TBM is  a full-mixing 
of $S_4$ but only a partial-mixing of $A_4$. The full-mixing of $A_4$ is given by the Cabibbo-Wolfenstein matrix
which gives too large a solar angle and too large a reactor angle. 
If $A_4$ is the correct symmetry group, as many models seem to prefer, one has to answer two questions.
(1) Why does Nature choose a partial mixing rather than a full-mixing more appropriate to $A_4$ from the symmetry (group-theoretical) point of view?
(2) Why should Nature choose a partial-mixing of $A_4$ and then  promotes it to a larger $S_4$ to give TBM, rather than choosing $S_4$
directly as the symmetry group from the very start?

With the result of Sec.~V, we may have an answer to question (1). If the Higgs self couplings are all positive, then the full-mixing pattern has a higher
energy than the partial-mixing pattern, which is presumably why Nature prefers partial mixing. Once in a partial-mixing mode, economy
would like to make the mixing matrix TBM because $\ol\G=S_4$ is the smallest group that contains $A_4$. 
Question (2) is harder to answer, but the answer might come from dynamics as well. In a forthcoming publication \cite{LAMNEW},
it will be shown that if we ignore the much smaller quark-mixing with its third generation, then there is something common between quark mixing and 
leptonic mixing  described by $A_4$, but there is no such commonality between quark mixing and leptonic mixing described by $S_4$.

\section{From Mixing to Horizontal Group}
We review in this section how the horizontal symmetry group $\G_L$ for left-handed leptons can be derived from the neutrino mixing matrix $U$ \cite{LAM}. We will assume $\G_L$ to be a subgroup of $SU(3)$, and we are particularly interested in small $\G_L$ to minimize the amount
of breaking necessary to obtain the desired mixing. For that reason we will concentrate on finite subgroups of $SU(3)$.

Let $M_e$ be the $3\x 3$ charged-lepton mass matrix, and $\ol M_e:=M_e^\dagger M_e$ the effective mass (squared) matrix connecting
left-handed  to left-handed charged leptons. This matrix is clearly hermitian. Similarly, let $\ol M_\nu=\ol M_\nu^T$
be the Majorana mass matrix for the left-handed neutrinos. This matrix is symmetric on account of the Majorana nature of the neutrinos.
In the basis where $\ol M_e$ is diagonal, the neutrino mixing matrix $U$
is the matrix that  renders $U^T\ol M_\nu U$ diagonal.

A {\it residual-symmetry operator} in the charged-lepton sector is a unitary matrix $F$ which transforms $\ol M_e$ into itself: 
$F^\dagger \ol M_e F=\ol M_e$. Similarly, a residual-symmetry operator in the neutrino sector is a unitary
matrix $G$ which transforms $\ol M_\nu$ into itself: $G^T\ol M_\nu G=\ol M_\nu$. $G^T$ rather than $G^\dagger$ has to be used to maintain
$\ol M_\nu=\ol M_\nu^T$. An important consequence that follows is $G^2=1$. As a result, the eigenvalues of $G$ are $+1$ or $-1$. Since
$\det G=1$, $G$ has one $+1$ and two $-1$ eigenvalues. There are exactly three such operators, given by
\be
G_1=u_1u_1^\dagger-u_2u_2^\dagger-u_3u_3^\dagger, \ G_2=-u_1u_1^\dagger+u_2u_2^\dagger-u_3u_3^\dagger, \ 
G_3=-u_1u_1^\dagger-u_2u_2^\dagger+u_3u_3^\dagger, \labels{G123}\ee 
where $u_i$ is the $i$th column vector of the mixing matrix $U$. It is easy to see that the three $G_i$'s mutually
commute, and that the product of any two is equal to the third. Moreover, if we define an {\it invariant eigenvector} to be
an eigenvector with eigenvalue $+1$, then $u_i$ is the invariant eigenvector of $G_i$.

Since $\ol M_e$ is diagonal, $F$ may be taken to be any diagonal unitary matrix, with determinant 1. If $F$ is to be a member of
a finite group, then there is an integer $n$ so that $F^n=1$. We divide $F$ into two categories, the degenerate ones, and the
{\it non-degenerate} ones. The latter consists of $F$'s whose three diagonal entries are all different, which necessitates
$n\ge 3$. This category is 
specially important because in that case $\ol M_e$ must be diagonal whenever $F$ is. This will be used in the next section to
recover $U$ from $\G_L$.

$F$ and $G_i$ are symmetry operators of the left-handed mass matrices, so they are the remaining symmetry operators of the {\it left-handed}
leptons after the horizontal group $\G_L$ for left-handed leptons is broken. Conversely, the smallest $\G_L$ must be the group
generated by $F$ and $G_i$, and this is how the group $\G_L$ can be obtained from the mixing matrix $U$. 
We shall use the notation
$\G_L=\{F,G_1,G_2,G_3\}$ to denote this. Since the product of two distinct $G_i$ is equal to the third, we can write
$\G_L=\{F,G_i,G_j\}$, as long as $i\not=j$. We shall call the group generated this way the {\it full group}.

The role of a non-degenerate $F$ is to ensure that $\ol M_e$ is diagonal when $F$ is, and the role of $G_i$ is to use their
invariant eigenvectors to construct the columns of the mixing matrix $U$. Since there is no need for more than one $F$ to nail down
$\ol M_e$ to its diagonal form, and since we want a minimal $\G_L$, we include only one $F$ in the generators of $\G_L$. However,
different choices of $F$ will generally lead to different groups $\G_L$.

Sometimes we may be more sure of one column of $U$ to be correct than another column. In that case we simply fill the other
two columns of $U$ with adjustable parameters, subject to the unitarity condition of $U$, of course. The full group $\ol \G$ 
associated with $U$ then contains a subgroup $\G_L$ generated by $\{F, G_i\}$, where
$i$ denotes that particular column of $U$ that we have confidence in. 
The group $\G_L$ will be called a {\it partial group}.
There are clearly three possible partial groups, associated with $i=1, 2,$ and 3 respectively.
If for the sake of economy we demand $\ol \G$ to be a finite group, then the
entries of the other two columns are severely restricted. In particular, if $\ol \G=\G_L$, then the partial group is a full group.

Note that since each $G_i$ is of order 2, a partial group must have an even order, and a full group must have an order divisible by 4.
Groups of odd order, such as $\Delta(27)$, would violate the assumption that at least one 
$G_i$ is a member of the unbroken horizontal group.

The symmetry of the right-handed
leptons cannot be determined by this procedure, because mixing has nothing to do with  right-handed fermions. Whenever it calls
for a knowledge of those, we shall make the minimal assumption that they are invariant under $\G_L$ as well. In other words,
if necessary, for simplicity we will implicitly assume $\G=\G_L$ in this paper.

For tribimaximal mixing \cite{HPS},
\be
U={1\over\sqrt{6}}\pmatrix{2&\rd&0\cr -1&\rd&\rt\cr -1&\rd&-\rt\cr}:=U_{0},\labels{TBM}\ee
the explicit form of $G_i$ using \eq{G123} works out to be
\be
G_{10}={1\over 3}\pmatrix{1&-2&-2\cr -2&-2&1\cr -2&1&-2},\ G_{20}={1\over 3}\pmatrix{-1&2&2\cr 2&-1&2\cr 2&2&-1},\ 
G_{30}=-\pmatrix{1&0&0\cr
	0&0&1\cr 0&1&0\cr}.\labels{GGG}\ee
To obtain a minimal group, we shall choose for $F$ the simplest non-degenerate matrix, $F_0=\diag(1,\o,\o^2)$,
where $\o=e^{2\pi i/3}$. In that case,
straight forward calculation using \eq{GGG} shows that the full group is $\G_L=\{F_0,G_{10},G_{20},G_{30}\}=S_4$, and the three partial
groups are $\{F_0,G_{10}\}=S_4, \{F_0,G_{20}\}=A_4, \{F_0,G_{30}\}=S_3$. 

Furthermore, it can be shown that as long as $F$ is non-degenerate, no matter what $n$ is ($F^n=1$),  the full group $\G_L$
always contain $S_4$ as a subgroup \cite{LAM}.

\section{From Horizontal Group to Mixing}
Given a finite group $\G_L$ containing a non-degenerate element $F$, we can reverse the argument of the last section to obtain 
all possible mixings associated with this group. Everything can be carried out in a purely group-theoretical manner, 
without  the presence of Higgs fields,
nor the help of a Lagrangian.

In doing so, we may assume $\G_L$ to be a full group, or a partial group. We will refer to the mixing matrices obtained with the former as
{\it full-mixing matrices} of $\G_L$, and those with the latter as {\it partial-mixing matrices}. Since partial-mixing matrices are
obtained from  full-mixing matrices  by replacing two columns with arbitrary entries, subject only to the unitarity constraint, there is 
no need to discuss them separately.

Another reason not to discuss them separately is that a partial-mixing matrix of $\G_L$ 
is a full-mixing matrix of a larger group $\ol\G$, as mentioned in the Introduction. Hence we need to study only full-mixing matrices.

The general procedure of constructing $U$ from $\G_L$ is as follows. 
First, separate the order-2 elements in $\G_L$ from those of order $\ge 3$. The former are candidates
of $G$, and the latter are candidates of $F$. Next. identify other order-2 elements $G'$ that commute with a $G$. Then $G''=GG'$ is an
order-2 element that commutes with both $G$ and $G'$, so the normalized
invariant eigenvectors  of $G, G', G''$ in the $F$-diagonal basis constitute the three columns of the full-mixing matrix $U$. 
Which eigenvector
occupies which column is a matter of convention, corresponding to different ways of labeling the three neutrino mass eigenstates,
or equivalently different ways of assigning $G, G', G''$ to be $G_1, G_2$, and $G_3$.
Moreover, the entries of the diagonal $F$ may be permuted, resulting in permutations of the rows of $U$. In other words, the full-mixing
matrix $U$ can be determined only up to permutation of rows and columns, because group theory has no way of knowing
which flavor neutrino is which, nor which mass neutrino eigenstate should be called 1, 2, or 3. When we present a $U$ below,
it is understood that such an ambiguity always exists.

We must check that $F,G,G',G''$ generate the group $\G_L$, and that 
the three normalized invariant eigenvectors in the $F$-diagonal basis are mutually orthogonal so that $U$ is unitary.

If we cannot find two order-2 elements in $\G_L$ that mutually commute, which would be the case for example when
the order of $\G_L$ is not divisible by 4, then $\G_L$ contains only partial-mixing matrices. The actual symmetry group
$\ol\G$ of the resulting mixing matrix must be larger, with an order divisible by four. 
 
Since Majorana phases are not measurable in neutrino oscillation experiments,
we will regard two $U$'s differing only by Majorana phases to be equivalent. With that we may classify neutrino mixing also by
the familiar Chau-Keung parametrization  used in quark mixing, which by convention leaves 
the (11), (12), (23), (33) matrix elements of $U$ real, through a suitable choice of the phases of the three columns
and the three rows. A mixing
matrix parametrized this way can be summarized by the triplet $\bk{\sin\th_{12}, \sin\th_{23}, \sin\th_{13}e^{-i\d}}$,
exhibiting the mixing angles $\th_{ij}$ and the CP-phase $\d$.
For TBM, this physical triplet is $\bk{{1\over\rt},{1\over\rd},0}$.

Although straight forward, the task of obtaining all the full-mixing matrices from a given group might seem  rather daunting
because of the large number of combinatorial choices of $F$ and $G$'s. In reality it is not as bad  because mixings
produced by members in the same conjugacy classes are often {\it equivalent}, {\it i.e.,}  they differ only by row and column phases, or the re-shuffling of rows and columns, so at least for relatively small groups $\G_L$, there are only very few inequivalent mixings. 

With such a foresight, we should arrange group elements into conjugacy classes.
One way to do so is to make use of the
character table, although characters other than those in the defining
three-dimensional representation will not be used in this section.
To illustrate the procedure, 
we shall discuss in the rest of this section how to obtain the full-mixing matrices from the groups $\G_L=S_4, A_4$,
and $A_5$. These are the rotational symmetry groups of the regular polyhedrons. 

The elements of each of these three groups can all
 be obtained by the repeated multiplication of two members,  $a$ and $b$, which are known as the
{\it generators}. In each case, $a$ obeys $a^2=1$ and $b$ obeys $b^3=1$. These three groups differ from one another in the behavior of $ab$,
which is of order 3, 4, 5 respectively for $A_4, S_4$, and $A_5$. In short, the {\it presentation} of these three groups are
\be
A_4&=&\bk{a,b|a^2=b^3=(ab)^3=1},\qquad S_4=\bk{a,b|a^2=b^3=(ab)^4=1},\nn\\
 A_5&=&\bk{a,b|a^2=b^3=(ab)^5=1}.\labels{pres}\ee

\subsection{$\bf S_4$}
$S_4$ is the group of permutation of four objects, with a  character table given in Table 1.
$$\ba{|c|ccccc|}\hline
&\C_1&\C_2&\C_3&\C_4&\C_5\\ \hline
&(1)&(12)&(12)(34)&(123)&(1234)\\ \hline
|{\cal C}_i|&1&6&3&8&6\\ \hline
\chi^1&1&1&1&1&1\\
\chi^2&1'&-1&1&1&-1\\
\chi^3&2&0&2&-1&0\\
\chi^4&3'&1&-1&0&-1\\
\chi^5&3&-1&-1&0&1\\
\hline
&&a&&b&ab\\ \hline \ea$$
\bc Table 1. Character table of $S_4$\ec
It contains five classes $\C_i\ (1=1,2,\cdots,5)$ with $|\C_i|$ elements,  defined by the cycle structure
of  permutations shown in the second row.  The order of the elements in each class can
be read off from the cycle structure to be 1, 2, 2, 3, 4, respectively.

There are five irreducible representations, $\chi^\a$,
with dimensions 1, 1, 2, 3, 3. The two 3-dimensional
irreducible representations IR4 and IR5 are very closely related. If $g^{3'}$ and $g^3$ are the representations of $g\in S_4$, 
respectively in IR4 and IR5, then $g^3=g^{3'}$ for $g\in \C_3$ or $\C_4$, and $g^3=-g^{3'}$ for $g\in\C_2$ or $\C_5$. The defining
three-dimensional representation (the matrices defining $F$ and $G_i$) is IR5, not IR4, because $\chi^5=-1$ but $\chi^4=+1$ for the class $\C_2$. Since
$G$ is of order 2,  it belongs to $\C_2$. It also has eigenvalues $+1, -1, -1$,
hence the trace of $G$, which is the character of $\C_2$, is $-1$ and not $+1$, so $G$ must belong to IR5. 

The last row lists where the generators $a$ and $b$ belong. From the order of the classes, clearly
$b$ has to belong to $\C_3$, but a priori $a$ could be in $\C_2$ or $\C_3$. 
$\C_3$ is excluded because then both $a$ and $b$ are even permutations, so the group they generate is the subgroup
$A_4$, not the whole $S_4$. Hence $a$ must be in $\C_2$.

Since the naming of the four objects under permutation is arbitrary, without
loss of generality we may take $b=(134)$. To be a generator of $S_4$ together with $b$, $a$ must then contain the number `2', so there are
three possibilities:  $a_1$=(12), $a_2=(23)=ba_1b^2$, and $a_3=(24)=b^2a_1b$. For each $a_i$, one can find an $a_i'$
and an $a_i''$ so that members of the triplet $A_i:=[a_i,a_i',a_i'']$ mutually commute and the product of two is the third. 
The explicit expression for the three triplets are
\be
A_1=[(12),(34),(12)(34)],\quad A_2=[(23), (14), (23)(14)],\quad A_3=[(24), (13), (24)(13)].\labels{triplets}\ee
Note that $a_i, a_i'\in\C_2$, but $a_i''\in\C_3$. 

Next, let us see how to represent the generators
as $SU(3)$ matrices, in the basis where $b=(134)$ is diagonal. Let us start off with $b$.
 Since $b^3=1$ and its character Tr($b$)=Tr($\C_4$) vanishes, its diagonal form must be $b=\diag(1,\o,\o^2)$
up to possible
permutation of the entries, where $\o=e^{2\pi i/3}$. This is the same as $F_0$ in TBM.
From \eq{GGG} and the discussion below it, we know that $G_{10}$ and $F_0$ together generate $S_4$, 
so we may take $G_{10}$ to represent any one of the generators $a_i$. For definiteness we let it be $a_1=(12)$.

Please note that notations such as $a, b$ may stand 
for the abstract elements defined by the permutation cycles, or matrices in various irreducible representations. 

To find the representation of the triplets $A_i$ in the $b$-diagonal basis, 
let $B:=(a_1b)^2=(14)(23)$. Then $Ba_1B^{-1}=a_1'$, which works out to be
$a_1'=G_{30}$ of \eq{GGG}. Furthermore, $a_1''=a_1a_1'=G_{20}$ of \eq{GGG}. Hence $A_1=[G_{10}, G_{30}, G_{20}]$.
Moreover, it is easy to check that $A_2=bA_1b^2$ and $A_3=b^2A_1b$. Hence $A_2=F_0[G_{10}, G_{30}, G_{20}]F_0^2$
and  $A_3=F_0^2[G_{10}, G_{30}, G_{20}]F_0$. 

Any one of the triplets $A_i$ is a possible choice for $\widetilde G:=[G,G',G'']$.
As to $F$, 
since $F$ is of order $\ge 3$, it is either in $\C_4$ or $\C_5$. Let us consider these two cases separately.

\bn
\i $\fbox{$F\in\C_4$}$.\quad In cycle notation, we can write $F=(xyz)$ for some numbers $x,y,z$ chosen among 1,2,3,4. The permutation
$g={\scriptsize \pmatrix{x&y&z\cr 1&3&4}}$  brings $F$ into diagonal form because $gFg^{-1}=(134)=b=F_0$. Since $F_0$ together with any
$A_i$ generate $S_4$, so does $F$ and $A_i^g:=g^{-1}A_ig$. The choice of $\widetilde G=(G,G',G'')$ would then be any of the three
$A_i^g$. 

To compute $U$, we need to go to the basis where $F$ is diagonal, in which case the representation of $\widetilde G$ is $A_i$
in the $b$-diagonal basis.
We may therefore assume from the outset that $F=b$ and $\tilde G=A_i$ for some $i$.

A similar step should be carried out for the cases to be considered later, but we will often skip that step and assume $F$ directly
 to be an appropriate generator of the group. 

\bn
\i \underline{$\widetilde G=A_1$}.\quad  If we let $\widetilde G=[G_1, G_3, G_2]$, then $G_i=G_{i0}$ and the 
full-mixing matrix is the TBM matrix $U_0$ in \eq{TBM}.

\i \underline{$\widetilde G=A_2$}. \quad  If we let $\widetilde G=[G_1, G_3, G_2]$, then $G_i=F_0G_{i0}F_0^2$
and the full-mixing matrix is just the TBM matrix $U_0$, multiplied respectively by $1, \o, \o^2$ in the first, second, and third rows.
Since row phases are adjustable, this is equivalent to the TBM in \eq{TBM}.

\i \underline{$\widetilde G=A_3$}. \quad  If we let $\widetilde G=[G_1, G_3, G_2]$, then $G_i=F_0^2G_{i0}F_0$
and the full-mixing matrix is just the TBM, multiplied respectively by $1, \o^2, \o$ in the first, second, and third rows.
Since row phases are adjustable, this is equivalent to the TBM in \eq{TBM}.

\i In conclusion, as long as $F\in\C_4$, the full-mixing matrix is equivalent to the TBM in \eq{TBM}, 
characterized by the physical parameters $\bk{{1\over\rt},{1\over\rd},0}$.
\en

\i $\fbox{$F\in \C_5$}$.\quad 
Any $F$ in this class obeys $F^{4}=1$, hence its possible eigenvalues are $1, +i, -i, -1$. Since we want it to be non-degenerate,
has determinant $+1$, 
    and since its character according to Table 1 is real, we conclude that its allowed eigenvalues are
	$1, i, -i$, each occurring once and only once. 
  For definiteness, we will let $F=a_1b=(1342):=c_1$. This is completely general because the four permutation objects in $S_4$
  can be assigned any label we wish to.

   Once again, $G$ must be chosen so that $\{F, G\}$ generate $S_4$. To see how to find such a $G$, write the presentation
	of $S_4$ in \eq{pres}  in a slightly different form, $S_4=\bk{a,c|a^2=c^4=(ac)^3=1}$, which shows that the generator
	$a$ is of order 2, the generator $c$ of of order 4, and $ac$
	is of order 3. Trying all 9 possibilities in $\C_2$ and $\C_3$ with $c=c_1=(1342)$, we see that only 
	the following four $a$'s, $a_1=(12), a_1'=(34),
	a_3'=(13),$ and $a_3= (24)$, 
	can yield an order-3 $ac$. Thus $G$ is confined to one of these four cases. As a consequence, the
	allowed triplets are $\widetilde G=A_1$ and $A_3$.

	In order to find out what these matrices are in the $F$-diagonal basis, we first write them  in the $b$-diagonal basis,
	then convert them by a similarity transformation to the $F$-diagonal bases.

	In the $b$-diagonal basis, 
\be
F=a_1b=G_{10}F_0={1\over 3}\pmatrix{1&-2\o&-2\o^2\cr -2&-2\o&\o^2\cr -2&\o&-2\o^2}.\labels{Fp}\ee
This $F$ can be diagonalized by the unitary matrix
\be
V={1\over\rt}\pmatrix{\o&-\o&\o\cr \o^2&-\o^2/(\rt+1)&-\o^2/(\rt-1)\cr 1&1/(\rt-1)&1/(\rt+1)\cr}\labels{VV}\ee
to yield $V^\dagger F V=\diag(1,i,-i):=F'$.

\bn
\i \underline{$\widetilde G=A_1$}.\quad In the $F$-diagonal representation, 
$[G_1,G_2,G_3]:=[G,G',G'']=V^\dagger [G_{10},G_{30},G_{20}]V$. The full-mixing matrix is then given by
\be U=V^\dagger U'_{0}=\pmatrix{-e^{\pi i/3}/\rd&-e^{-\pi i/6}/\rd&0\cr 
-e^{-5\pi i/12}/2&-e^{\pi i/12}/2&e^{\pi i/12}/\rd\cr-e^{\pi i/12}/2&e^{-5\pi i/12}/2&e^{-5\pi i/12}/\rd\cr},\labels{uc5}\ee
where $U'_{0}$ is the TBM mixing matrix given in \eq{TBM} with the second and third columns switched. We do that switching
 to put $U$ into a more familiar form with a vanishing reactor angle.

Multiplying the first, second, and third columns by $e^{\pi i/6}, e^{-\pi i/12}, e^{5\pi i/12}$, and the first 
and second columns by $i, -1$, we
can turn $U$ into the equivalent form
\be
U\to \pmatrix{1/\rd&1/\rd&0\cr -1/2&1/2&1/\rd\cr 1/2&-1/2&1/\rd\cr},\labels{uc51}\ee
whose mixing angles are given by the physical triplet $\bk{{1\over\rd},{1\over\rd},0}$.
In this case, the reactor angle remains to be zero, but both the solar and the atmospheric mixings are maximal.

\i \underline{$\widetilde G= A_3$}.\quad 
First, note that $A_3=[(24), (13), (24)(13)]=c_1[(34),(12),(34)(12)]c_1^{-1}$, where $c_1=(1342)$. In the $b$-diagonal representation,
$[a_3', a_3, a_3'']=c_1[a_1,a_1',a_1'']c_1^{-1}=F[G_{10},G_{30},G_{20}]F^{-1} $. In the $F$-diagonal basis, 
$[G_1,G_2,G_3]:=[G', G, G'']=V^\dagger F[G_{10},G_{30},G_{20}]F^{-1} V=F'V^\dagger[G_{10}, G_{30}, G_{20}]VF^{'-1}$. The full-mixing
matrix is therefore $F'V^\dagger U'_{TBM}=F'U_5$, with $U_5$ being the $U$ in \eq{uc5}. Since $F'$ is a diagonal phase matrix,
the full-mixing matrix here is equivalent to \eq{uc5} and \eq{uc51}.

\i   In conclusion, as long as $F\in\C_5$, the full-mixing matrix is equivalent to $U$ of \eq{uc51}, characterized by the physical parameters $\bk{{1\over\rd},{1\over\rd},0}$.
\en

\en
To summarize, up to equivalence, 
the full-mixing matrix of $S_4$ is either given by the TBM in \eq{TBM}, characterized by the physical parameters
$\bk{{1\over\rt},{1\over\rd},0}$, or the matrix \eq{uc51}, characterized by $\bk{{1\over\rd},{1\over\rd},0}$.

\vss
\subsection{$\bf A_4$}
The character table of $A_4$ is given in Table 2.
$$\ba{|c|cccc|}\hline
&\C_1&\C_2&\C_3&\C_4\\ \hline
&(1)&(12)(34)&(123)&(132)\\ \hline
|{\cal C}_i|&1&3&4&4\\ \hline
\chi^1&1&1&1&1\\
\chi^2&1'&1&\o&\o^2\\
\chi^3&1''&1&\o^2&\o\\
\chi^4&3&-1&0&0\\
\hline
&&a&b&b^2\\ \hline \ea$$
\bc Table 2. Character table of $A_4$\ec
In this case the order-3 elements are divided into two classes, $\C_3$ and $\C_4$,
with the elements in $\C_4$ being the square, or the inverse, of the elements in $\C_3$.
Since the generator $b$ is of order 3, it is either in $\C_3$ or $\C_4$, which differ
only by the arbitrary naming of the four objects. For definiteness we will choose it
to be in $\C_3$. Since its three dimensional character $\chi^4$ vanishes, up to permutation
of the entries, the diagonal form of $b$ is again $b=\diag(1,\o,\o^2)$. For definiteness,
once again we will let $b=(134)$.

The generator $a$ is of order 2, so it must be in $\C_2$. To be a generator, it must contain
the permutation of the object `2', but then all of them do. Hence there are three possibilities: 
$a_1=(12)(34),\ a_2=(13)(24),\ a_3=(14)(23)$. These three mutually commute, and the product
of any two is the third, so they can be taken to be $G, G', G''$ respectively.

In the diagonal form, a non-degenerate $F$ must be $F=b=\diag(1,\o,\o^2)=F_0$, up to permutation of its entries.
We must now find out the representation of $A=(a_1,a_2.a_3)$ in this basis.

We know from \eq{GGG} and the discussions below that equation that $F_0$ and $G_{20}$ generate $A_4$, hence we may take 
$a_2=G_{20}$. Then $a_1=(12)(34)=(134)(13)(24)(143)=ba_1b^2:=G_1'$, and $a_3=(14)(23)=b^2a_1b:=G_3'$. 
Since the invariant eigenvector of $G_2$ is proportional
to $(1,1,1)^T$, and invariant eigenvectors of $G_1'$ and $G_3'$ are respectively $F_0(1,1,1)^T=(1,\o,\o^2)^T$ and 
$F_0^2(1,1,1)^T=(1,\o^2,\o)^T$. Consequently, the full-mixing matrix of $A_4$ is the Cabibbo-Wolfenstein mixing matrix
\be
U={1\over\rt}\pmatrix{1&1&1\cr \o&1&\o^2\cr \o^2&1&\o\cr},\labels{cw}\ee
whose triplet of physical parameters can be shown to be $\bk{{1\over\rd},{1\over\rd},{e^{-i\pi/2}\over\rt}}$.

\vss
\subsection{$\bf A_5$}
The character table of $A_5$ is given in Table 3,
$$\ba{|c|ccccc|}\hline
&\C_1&\C_2&\C_3&\C_4&\C_5\\ \hline
&(1)&(12)(34)&(123)&(13524)&(12345)\\ \hline
|{\cal C}_i|&1&15&20&12&12\\ \hline
\chi^1&1&1&1&1&1\\
\chi^2&4&0&1&-1&-1\\
\chi^3&5&1&-1&0&0\\
\chi^4&3'&-1&0&-1/\varphi&\varphi\\
\chi^5&3&-1&0&\varphi&-1/\varphi\\
\hline
&&a&b&ab&(ab)^2\\ \hline \ea$$
\bc Table 3. Character table of $A_5$\ec
where $\varphi$ is the golden ratio $(1+\sqrt{5})/2=-2\cos(4\pi/5)=1.618$, and 
$1/\varphi=(-1+\sqrt{5})/2=\varphi-1=2\cos(2\pi/5)=0.618$.

The elements in class $\C_5$ are the square of the elements in $\C_4$. From the presentation
of the group given in \eq{pres}, we see that the generators $a$
and $b$ must in classes $\C_2$ and $\C_3$ respectively, and $ab$ must in $\C_4$ or $\C_5$. Which
of the two it belongs simply depends on how we label the five objects in $A_5$.

For definiteness, let $b=(134)$ as before. To be a generator of $A_5$, $a$ must contain the numbers `2' and `5'
in separate cycles, 
such as $a_{13}=(12)(35)$ or $a'_{13}=(15)(23)$. These two commute and produce a product  $a''_{13}=(13)(25)$,
which incidentally is not a generator because `2' and `5' are in the same cycle. In addition to $A_{13}=[a_{13},a'_{13},a''_{13}]$,
there are two other commuting triplets, $A_{14}=[a_{14},a'_{14},a''_{14}]$ and $A_{34}=[a_{34},a'_{34},a''_{34}]$,
with $a_{14}=(12)(45), a'_{14}=(15)(24), a''_{14}=(14)(25)$,
 and $a_{34}=(32)(45), a'_{34}=(35)(24), a''_{34}=(34)(25)$. In each case, $a$ or $a'$ is a possible generator of $A_5$, but not
$a''$. Any of these three triplets is a possible choice for $\widetilde G=[G, G', G'']$.

$F$ may be in $\C_3$ or $\C_4$, but we will only consider the case $F\in\C_3$ here 
\footnote{ There have been some model calculations of $A_5$ \cite{A5}, which agree with the group-theoretical result given below.
 I am grateful to Prof. Feruglio for informing me that the result of the case not considered here is also contained in their paper.}
. As usual, without loss of generality we may choose
$F=b=(134)=\diag(1,\o,\o^2)=F_0$ as before. The question is what is the three-dimensional representation for the generator $a$.

Representation of $A_5$ can be found in \cite{A5rep}, but unfortunately it is not in the $b$-diagonal basis that we need. 
Nevertheless, from that 
representation, one can work out the defining representation $a$ in the $b$-diagonal basis to be 
\be
a={\footnotesize
\pmatrix{{\rc\over 3}&-{1\over 4}+{1\over 12}(\rc+i\rt+i\rc\rt)&-{1\over 4}+{1\over 12}(\rc-i\rt-i\rc\rt)\cr
-{1\over 4}+{1\over 12}(\rc-i\rt-i\rc\rt)&-{1\over 2}-{\rc\over 6}&{1\over 4}+{1\over 12}(-2\rc-2i\rt+i\rc\rt)\cr
-{1\over 4}+{1\over 12}(\rc+i\rt+i\rc\rt)&{1\over 4}+{1\over 12}(-2\rc+2i\rt-i\rc\rt)&-{1\over 2}-{\rc\over 6}\cr}}.\nn\\
\labels{aa5}\ee
One can check that $b^3=a^2=1$ and $(ab)^5=1$, so these $a$ and $b$ are indeed generators of $A_5$.

This $a$ can be equated to any one of the generators in $\C_2$. For definiteness, let it be $a=a_{13}=(12)(35)$. We must also
find the matrix representation of its commuting partners $a'_{13}:=a'$ and  $a''_{13}:=a''$. They are in the same class as $a$, 
so each of them can be 
written in the form $gag^{-1}$, for some $g$ which itself a product of the generators $a$ and $b$. It can be shown that $g=(ba)^2b^2=(125)$
and $g^{-1}=b(ba)^3=(152)$ convert $a=(12)(35)$ into $a''=(13)(25)$. Using the representation of $b$ and $a$, one can calculate from \eq{aa5} to obtain
\be
a''={\footnotesize{1\over 8}\pmatrix{-8&0&0\cr 0&0&1+3\rc+\rt i(1-\rc)\cr 0&1+3\rc-\rt i(1-\rc)&0\cr}},\labels{a3a5}\ee
and then from $a'=aa''$ that 
\be
a'=\footnotesize{
\pmatrix{-{\rc\over 3}&{1\over 4}-{1\over 12}(\rc+i\rt+i\rc\rt)&{1\over 4}-{1\over 12}(\rc-i\rt-i\rc\rt)\cr
{1\over 4}-{1\over 12}(\rc-i\rt-i\rc\rt)&-{1\over 2}+{\rc\over 6}&{1\over 4}+{1\over 12}(-2\rc-2i\rt+i\rc\rt)\cr
{1\over 4}-{1\over 12}(\rc+i\rt+i\rc\rt)&{1\over 4}+{1\over 12}(-2\rc-2i\rt-i\rc\rt)&-{1\over 2}+{\rc\over 6}\cr}}.\nn\\
\labels{a2a5}\ee

The unitary matrix to diagonalize simultaneously $a, a', a''$ so that $U^\dagger a U=\diag(+1,-1,-1),\
U^\dagger a' U=\diag(-1,+1,-1),\ U^\dagger a'' U=\diag(-1,-1,+1)$ is
\be U={\footnotesize{1\over\rt}\pmatrix{\varphi&\varphi^{-1}&0\cr
-{1\over 2}(1+\o\varphi)\varphi^{-1}&{1\over 2}(1+\o\varphi)\varphi&{\rt\over 2}i(1+\o\varphi)\cr
{1\over 2}(1+\o^2\varphi)\varphi^{-1}&-{1\over 2}(1+\o^2\varphi)\varphi&{\rt\over 2}i(1+\o^2\varphi)\cr}}.\labels{ua5}\ee
Writing $(1+\o\varphi)=(1+\o^2\varphi)^*=\rd e^{i\xi}$, then multiply the second and the third rows of $U$ by $e^{-i\xi}$ and 
$-e^{i\xi}$ respectively, and the third column by $-i$, we can turn $U$ into the equivalent form
\be
U\to {1\over\rt}\pmatrix{\varphi&\varphi^{-1}&0\cr -{1\over\rd\varphi}& {\varphi\over\rd}&{\rt\over\rd}\cr 
{1\over\rd\varphi}& -{\varphi\over\rd}&{\rt\over\rd}\cr},\labels{ua52}\ee
whose physical parameters are given by the triplet $\bk{{1\over\rt\varphi},{1\over\rd},0}$. This is the full-mixing
matrix of $A_5$ if $(G, G', G'')=(a_{13}, a'_{13}, a''_{13})=(a,a',a'')=((12)(35), (15)(23), (13)(25))$.

Next, consider the other triplet $(G, G', G'')=(a_{14}, a'_{14}, a''_{14}) $=$((12)(45), (15)(24),(14)(25)) $. 
Let $g=(ba)^3=((134)(12)(35))^3=(15243)$. Then $(a_{14}, a'_{14}, a''_{14})=g(a,a'',a')g^{-1}$. Note that the positions of
$a''$ and $a'$ are reversed. The full-mixing matrix for this triplet choice is therefore $U=X\tilde U$, where
$\tilde U$ is the matrix in \eq{ua5} with the second and third columns interchanged, and 
\be
X=(ba)^3={\footnotesize{1\over 12}\pmatrix{4&1-i\rt+\rc(3+i\rt)&1+i\rt+\rc(3-i\rt)\cr
4(1-i\rt)&1+i\rt-\rc(3-i\rt)&1-i\rc\rt\cr
4(1+i\rt)&1+i\rc\rt&1-i\rt-\rc(3+i\rt)\cr}}.\labels{X}\ee
The result is
\be
U=X\tilde U={\footnotesize{1\over \rt}\pmatrix{\varphi^{-1}&-\varphi&0\cr
{1\over\rd}\varphi e^{-i\eta}&{1\over\rd}\varphi^{-1} e^{-i\eta}&-i{\rt\over\rd}e^{-i\eta}\cr
{1\over\rd}\varphi e^{i\eta}&{1\over\rd}\varphi^{-1} e^{i\eta}&i{\rt\over\rd}e^{-i\eta}\cr
}},\labels{ua53}\ee
where $\eta$ is defined by ${\sqrt{15}\over 12}\pm {i\over 4}={1\over\rs}e^{\pm i\eta}$.

Multiply the second row by $e^{i\eta}$, the third row by $-e^{-i\eta}$, the second column by $-1$, the third column by $i$,
and interchange the first and second columns, we get back to the mixing in \eq{ua52}. 

Finally, similar calculation shows that if $(G,G',G'')=(a_{34}, a'_{34}, a''_{34})$, then its full-mixing matrix is also equivalent
to \eq{ua5}.

In summary, as long as $F\in C_3$, the full-mixing matrix of $A_5$ is equivalent to \eq{ua52}, a mixing which is characterized by the
triplet physical mixing parameters 
$\bk{{1\over\rt\varphi},{1\over\rd},0}$.

\section{Connection between Group Theory and Dynamics}
Charged-lepton masses come from Yukawa interactions in the Standard Model. If  neutrino masses arise from the type-I
seesaw mechanism, then the Lagrangian responsible for flavor structure of mass matrices can be symbolically written as
\be
{\cal L}=-c_\g \ol e_R e_L\f^\g-h'_\g \ol N_R\nu_L\chi^{'\g}-{1\over 2}h_\g N_R^TN_R \chi^{\g}+h.c.,\labels{L1}\ee
where $e_L, e_R$ are the left-handed and right-handed charged leptons,  $\nu_L$, $N_R$ are the active and heavy Majorana neutrinos,
and $\f^\g, \chi^\g, \chi^{'\g}$ are the flavons.
Isotopic spin and spacetime details are omitted, so this Lagrangian should only be used  to discuss the flavor structure of mass
matrices.
In that regard every term is assumed to be invariant under a horizontal symmetry group $\G$, with $c_\g, h_\g, h'_\g$ being
the Yukawa coupling constants. The index $\g$ labels the irreducible representations (IR) of $\G$, and a sum over all $\g$ is understood. 
The IR of the fermions are  not explicitly specified.

For type-II seesaw, we replace the last two terms with a single term, hence
\be
{\cal L}=-c_\g \ol e_R e_L\f^\g-h_\g \nu_L^T\nu_L\chi^\g+h.c.\labels{L2}\ee

To obtain the mass matrices from \eq{L1} or \eq{L2}, vacuum expectation values of the flavons are introduced to break the $\G$-symmetry. In order for the residual symmetries to be preserved after the breaking, $F$ in the charged-lepton sector and $G_i$ in the neutrino sector, the vacuum alignments have to be invariant eigenvectors (eigenvector with eigenvalue $+1$) of
the corresponding residual operators \cite{LAM},
\be
F^{(\g)}\bk{\f^\g}=\bk{\f^\g},\quad G^{(\g)}_i\bk{\chi^\g}=\bk{\chi^\g}, \quad G^{(\g)}_i\bk{\chi^{'\g}}=\bk{\chi^{'\g}}, \labels{inveig}\ee
where $F^{(\g)}$ and $G^{(\g)}_i$ are the representations of $F, G_i$ in IR$\g$. If we want to obtain a full-mixing matrix, then
\eq{inveig} is required to be obeyed for $i=1,2,3$, but if we want to construct a partial-mixing matrix, then the equation has
to be satisfied only for that $i$.

Assuming the invariant eigenvectors to be unique, it follows from \eq{inveig} that $\bk{\chi^\g}=\bk{\chi^{'\g}}$
up to normalization, so we will set $\chi^{'\g}=\chi^\g$ in \eq{L1} from now on.

Note that \eq{inveig} does not rely on the specific form of \eq{L1} or \eq{L2}. All that it requires is   
the neutrino mass terms to be distinct from the charged-lepton mass terms, and that each has its own valon. 
\eq{inveig} would be equally valid if valon fields appear quadratically in the Lagrangian, for example.

Our main goal is to keep the effective left-handed mass matrices $\ol M_e$ and $\ol M_\nu$ invariant under the residual symmetry.
This can be accomplished by keeping the broken Lagrangian invariant, as is done above, but in the case of type-I seesaw, since
the Dirac mass matrix $M_\nu$ enters quadratically in $\ol M_\nu=M_\nu^TM_N^{-1}M_\nu$, 
where $M_N$ is the Majorana mass matrix of $N_R$, invariance of $\ol M_\nu$
can also be preserved {\it indirectly} \cite{KING} if the last term in \eq{inveig} is replaced by $G^{(\b)}_i\bk{\chi^{'\b}}
=-\bk{\chi^{'\b}}$.

Once we know the IR of the residual symmetry operators, vacuum alignments can be calculated. With those alignments,  one can 
construct many models
that automatically give rise to
the desired neutrino mixing.  Yukawa coupling constants  are used to fit the
fermion masses, and in the case
of partial mixing, also the remaining mixing parameters.
Different models are obtained by choosing different
IR assignments of $e_L, e_R, \nu_L, N_R, \phi$, and $\chi$, though for a successful model we must make sure that enough terms are present
in each sector to fit all the masses and the remaining mixing parameters, in spite of the zero vacuum alignments
forced upon us in those IR without an invariant eigenvector.

Let us see how this works explicitly. Let $\bk{\a a,\b b|\g c}$ be the Clebsch-Gordan (CG) coefficient of the group $\G$, coupling
orthonormal states $\ket{\a a}$ in  IR$\a$ with the orthonormal states $\ket{\b b}$ in IR$\b$ to get an orthonormal state
$\ket{\g c}$ in  IR$\g$,
\be \ket{\g c}&=&\sum_{a,b}\ket{\a a}\ket{\b b}\bk{\a a,\b b|\g c},\nn\\
\ket{\a a}\ket{\b b}&=&\sum_{\g,c}\ket{\g c}\bk{\g c|\a a,\b b}.
\labels{CGC}\ee
The CG coefficients must obey the group-invariant condition
\be
\sum_{c'}\bk{\a a,\b b|\g c'}g^{(\g)}_{c'c}&=&\sum_{a',b'}g^{(\a)}_{aa'}g^{(\b)}_{bb'}\bk{\a a',\b b'|\g g},\nn\\
\sum_{c'}g^{(\g)}_{cc'}\bk{\g c'|\a a,\b b}&=&\sum_{a',b'}\bk{\g c|\a a',\b b'}g^{(\a)}_{a'a}g^{(\b)}_{b'b}.
\labels{CGI}\ee
for every $g\in\G$, or at least for every generator of $\G$. It is important to note that if we change the IR bases of $g$,  
the CG coefficients may have to be
changed as well. In the case of $S_4$, the CG coefficients to be used will be those that correspond to the IR given in Table 4.

Suppose $e_R$ and $e_L$ belong to IR$\a$ and IR$\b$ respectively, and the flavons belong to IR$\g$. Then the charged-lepton
mass matrix can be read out from either \eq{L1} or
\eq{L2} to be 
\be (M_e)_{ab}=\sum_{\g,c}c_\g\bk{\a^* a, \b b|\g c}\bk{\f^\g_c}.\labels{me}\ee
The invariance of the charged-lepton mass matrices $M_e$ and  $\ol M_e:=M_e^\dagger M_e$ under the residual symmetry $F$
can be seen from \eq{inveig} and \eq{CGI}:
\be F^{(\a) \dagger}M_eF^{(\b)}=M_e,\quad
F^{(\b)^\dagger}\ol M_e F^{(\b)}=\ol M_e.\labels{finv}\ee 

Similarly, if $N_R, \nu_L$ belong to IR$\a$ and IR$\b$ respectively, then the neutrino mass matrices are
\be (M_\nu)_{ab}=\sum_{\g,c}h'_\g\bk{\a a,\b b|\g c}\bk{\chi^\g_c}, \ (M_N)_{aa'}=\sum_{\g,c}h_\g\bk{\a a,\a a'|\g c}_S\bk{\chi^\g_c}, \
 \ol M_\nu=M_\nu^TM_N^{-1}M_\nu,\labels{mnu}\ee
where $\bk{\a a,\a a'|\g c}_S$ is the CG coefficient symmetric in $a$ and $a'$. Once again the invariance relations
\be 
G_i^{(\a)\dagger}M_\nu G_i^{(\b)}=M_\nu,\ G_i^{(\a)\dagger}M_NG_i^{(\a)*}=M_N,\ G_i^{(\b)T}\ol M_\nu G_i^{(\b)}=\ol M_\nu.
\labels{giinv}\ee
follow. Note that because of the Majorana nature of the neutrino, we have assumed $G_i^{(\a,\b)}$ to be real. For indirect models, the first 
equation in \eq{giinv} should have a minus sign.

In the rest of this section, we will illustrate these operations by considering
the direct full-mixing models of the groups $\G=A_4$ and $S_4$.
In the case of $A_4$, partial-mixing models are also discussed because they can give give rise to the phenomenologically interesting
tribimaximal mixing.

The first task in each case is to find out the IR of the residual symmetry operators.

\vss
\subsection{$\bf S_4$}
Remember from Sec.~IV that there are two full-mixing modes, one giving rise to the TBM, and the other yields a mixing with too large a solar angle.
We shall consider how dynamical models can be constructed in each case, starting with those giving rise to the TBM.
 
\bn
\i $\fbox{$F=F_0$,\quad $[G_1,G_2,G_3]=[G_{10}, G_{20}, G_{30}]$}$

Table 4 shows the IR and invariant eigenvectors of the residual symmetry operators in the basis where $F$ is diagonal.

$$\ba{|c|c|c|c|c|c|}\hline
{\rm IR}&{\bf 1}&{\bf 1'}&{\bf 2}&{\bf 3'}&{\bf 3}\cr\hline\hline
F&1&1&\diag(\o,\o^2)&\diag(1,\o,\o^2)&\diag(1,\o,\o^2)\cr\hline
\bk{\f}&1&1&(0,0)&(1,0,0)&(1,0,0)\cr\hline\hline
G_1&1&-1&\sigma_1&-G_{10}&G_{10}\cr\hline
G_2&1&1&\diag(1,1)&G_{20}&G_{20}\cr\hline
G_3&1&-1&\sigma_1&-G_{30}&G_{30}\cr\hline
\bk{\chi}&1&0&(1,1)&(1,1,1)&(0,0,0)\cr
\hline\ea$$

\vss
\bc Table 4. Irreducible representations and invariant eigenvectors of the first set of $S_4$ residual symmetry operators\ec

The one-dimensional IR ${\bf 1,\ 1'}$ in Table 4 are obtained directly from Table 1 because in that case character is the IR.
The three-dimensional representations ${\bf 3,\ 3'}$ are copied from those given in the previous section.
For {\bf 2}, 
since $F^3=1$ and its character from Table 1 is ${\rm Tr}(\C_4)=-1$, its diagonal form must be $\diag(\o,\o^2)$. The other generator $G_1$
must not be diagonal, or else the 2-dimensional representation is not irreducible. Since $G_1^2=1$, within equivalence we
may choose it to be the Pauli matrix ${\scriptsize \sigma_1=\pmatrix{0&1\cr 1&0\cr}}$.

With these IR of the residual symmetry operators, it is straight forward to work out the invariant eigenvector $\bk{\f}$
of $F$, and the simultaneous invariant eigenvector $\bk{\chi}$ of $G_1, G_2$, and $G_3$. The column vectors are displayed
in the table as row vectors for printing convenience. Some of the vacuum alignments
are zero either because  $F^\a$ does not have $+1$ eigenvalue, or because there is no common eigenvector of $G_1^\a,G_2^\a, G_2^\a$
all with eigenvalue $+1$.

With the help of Tables 1 and 4,  many  direct models 
possessing TBM  can be constructed. These models differ from one another
in their IR assignments for the right-handed fermions and flavons.

We saw in the last section that the residual symmetry operators for $S_4$ and $A_4$ are in IR{\bf 3}, hence the left-handed
fermions must transform as {\bf 3}. One can then  work out from the character table the allowed IR of the flavons for every
 assignment of the right-handed fermion. The result  is listed in row 2 of Table 5. Since $\bk{\f}=0$ for 
IR{\bf 2} and $\bk{\chi}=0$ for IR{\bf 3},  neither of them can be used to construct models. Taking that into account,
the final number of free parameters available in model construction
is listed in the third row for the charged-lepton sector, the fourth row for
the neutrino sector if it is given by a type-II seesaw, and the fifth row for the neutrino sector given by a type-I seesaw.
Those in the neutrino sector require a little bit of explanation.

For type-II seesaw, $\nu_L$ is coupled to itself so only ${\bf 3}\x{\bf 3}$ coupling is relevant. For type-I seesaw, the heavy Majorana neutrino $N_R$ can be assigned
to any IR. The first of the two numbers in the last row refers to the number of free parameters in the
Dirac neutrino mass term. Once the assignment ${\bf \a}$ of $N_R$ is determined, the number of parameters in the Majorana
mass terms is determined by $\a\x\a$, and that appears as the second number in the last row. 
If we consider indirect models \cite{KING}, then the number of parameters is even larger than that.

$$\ba{|c|c|c|c|c|c|}\hline
{\rm RHF}&{\bf 1}&{\bf 1'}&{\bf 2}&{\bf 3'}&{\bf 3}\cr\hline
{\rm flavons}&{\bf 3}&{\bf 3'}&{\bf 3', 3}&{\bf 1', 2, 3', 3}&{\bf 1, 2, 3', 3}\cr\hline\hline
e\ {\rm param}&1&1&2&3&3\cr\hline
{\rm typeII}\ \nu&-&-&-&-&3\cr\hline
{\rm typeI}\ \nu&0+1&1+1&1+3&3+3&3+3\cr
\hline\ea$$

\bc Table 5. Number of free parameters in $S_4$ models with a TBM mixing\ec

Let us see what Table 5 is telling us. First look at the charged-lepton sector, in row 3. Since the number free parameters always happens
to match the dimension of the representation, there are always just enough of them to fit the three charged-lepton masses. 

For type-II seesaw, there is again exactly three free parameters to fit the three active neutrino masses. For type-I seesaw,
the number of free parameters shown is the sum of the $M_\nu$ and $M_N$ parameters, which is generally more than enough 
to fit the active neutrino masses. The remaining ones can be used, for example, to tune the properties of the heavy neutrinos.

As a concrete illustration, we will write down the mass matrices of
 a type-I model where $e_L, e_R, \nu_L, N_R$ all belong to {\bf 3}. It is the kind of assignment that a
$SO(10)$ grand-unified theory would want. In that case, according to Tables 4 and  5, $M_e$, $M_\nu$, and $M_N$ should contain
exactly three parameters each. Using the CG coefficients in the $F_0$-diagonal basis, these mass matrices can be worked out to be
$M_e=\diag(\ol c_1+2\ol c_{3'},\ \ol c_1-\ol c_{3'}+\ol c_3,\ \ol c_1-\ol c_{3'}-\ol c_3)$, where $\ol c_1=c_1/\rt,\ \ol c_{3'}=c_{3'}/3\rd$,
and $\ol c_3=c_3/\sqrt{6}$. Furthermore,
\be
M_N=\pmatrix{\ol h_1+2\ol h_{3'}&\ol h_2-\ol h_{3'}&\ol h_2-\ol h_{3'}\cr \ol h_2-\ol h_{3'}&\ol h_2-\ol h_{3'}&\ol h_1+\ol h_{3'}\cr \ol h_2-\ol h_{3'}& \ol h_1+\ol h_{3'}&\ol h_2-\ol h_{3'}\cr},\labels{s4mn}\ee
where $\ol h_1=h_1/\rt,\ \ol h_{3'}=h_{3'}/3\rd,$ and  $\ol h_2=h_2/\sqrt{6}$. The expression for $M_\nu$ is identical except that $h_\a$ is to be replaced by
$h'_\a$.

This example illustrates some general points discussed above:
\bn
\i Each of $M_e, M_\nu, M_N$ is specified by three independent parameters, as shown in Table 5 above.
\i The three parameters in $M_e$ can be fitted by the three charged-lepton masses.
\i There are six parameters to specify $\ol M_\nu=M_\nu^TM_N^{-1}M_\nu$. In additional to fitting the three active 
neutrino masses, we can use
the remaining three to specify the heavy $N_R$ masses.
\i $M_e$ is automatically diagonal. This is because we are using CG coefficients in the $F_0$-diagonal (hence $\ol M_e$-diagonal) basis. 
As remarked below \eq{CGI}, if we use some other basis, then the CG coefficient will be rotated and the resulting $M_e$ may no longer be diagonal.
\i Whatever the parameters are, both $M_\nu$ and $M_N$ turn out to be magic (all row sums are equal) and 2-3 symmetric (invariant under
a simultaneous exchange of the second and third columns, and second and third rows). Hence $\ol M_\nu=M_\nu^TM_N^{-1}M_\nu$ is also magic and
2-3 symmetric, and
the mixing matrix $U$ is 
automatically TBM \cite{23magic}.
\en

\vskip1cm

\i $\fbox{$F=G_{10}F_0$,\quad $[G_1,G_2,G_3]=[G_{10}, G_{20}, G_{30}]$}$

This is the other allowed full mixing of $S_4$, given in \eq{Fp}, \eq{VV}, \eq{uc5}, and \eq{uc51}. Unlike the TBM mixing, whose
models abound in the literature, this one is seldom discussed presumably because its phenomenology is less attractive.
Another reason might be that the appropriate vacuum alignments in this case are a bit strange, as we shall see. 
Nevertheless, we can follow the general procedure to construct models that lead to this kind of mixing. 

There are two different ways to proceed. Either we go directly to the $F$-diagonal basis, as prescribed before, or we stay in the
$F_0$-diagonal basis that we are familiar with. The former has the advantage of having $\ol M_e$ automatically diagonal, but the CG coefficients in the $F$-diagonal basis must be worked out anew. If we travel the latter route, the CG coefficients are the same
ones used in \eq{s4mn}, but we have to work out the new alignments, and have to perform the diagonalization of $\ol M_e$
afterward. To illustrate better the connection between the group-theoretical and the dynamical approaches, we choose to work with the latter.

In the $b$-diagonal basis, the IR of $G_i$ is the same as in Table 4, and the
IR of $F$ can be obtained from Table 4 by calculating $G_1F$. The resulting IR as well as the
invariant eigenvectors are shown in Table 6 below. Since the IR for $G_i$ remains unchanged, that part of the table is identical to Table 4.
The only difference comes in the upper part of the table in which $F$ is replaced by $G_{1}F$ in every IR.

\vss
$$\ba{|c|c|c|c|c|c|}\hline
{\rm IR}&{\bf 1}&{\bf 1'}&{\bf 2}&{\bf 3'}&{\bf 3}\cr\hline\hline
F&1&-1&\sigma_1\diag(\o,\o^2)&-G_{10}F_0&G_{10}F_0\cr\hline
\bk{\f}&1&0&(1,\o)&(0,0,0)&(\o,\o^2,1)\cr\hline\hline
G_1&1&-1&\sigma_1&- G_{10}& G_{10}\cr\hline
G_2&1&1&\diag(1,1)& G_{20}&G_{20}\cr\hline
G_3&1&-1&\sigma_1&- G_{30}& G_{30}\cr\hline
\bk{\chi}&1&0&(1,1)&(1,1,1)&(0,0,0)\cr
\hline\ea$$

\vss
\bc Table 6. Irreducible representations and invariant eigenvectors of the second set of $S_4$ residual symmetry operators\ec

\vss

The parameter count for this kind of models is shown in Table 7. The only variation  from Table 5 comes from the different vanishing $\bk{\f}$'s, now occurring in ${\bf 1'}$ and ${\bf 3'}$ instead of {\bf 2}. 

$$\ba{|c|c|c|c|c|c|}\hline
{\rm RHF}&{\bf 1}&{\bf 1'}&{\bf 2}&{\bf 3'}&{\bf 3}\cr\hline
{\rm flavons}&{\bf 3}&{\bf 3'}&{\bf 3', 3}&{\bf 1', 2, 3', 3}&{\bf 1, 2, 3', 3}\cr\hline\hline
e\ {\rm param}&1&0&1&2&3\cr\hline
{\rm typeII}\ \nu&-&-&-&-&3\cr\hline
{\rm typeI}\ \nu&0+1&1+1&1+3&3+3&3+3\cr
\hline\ea$$

\bc Table 7. Number of free parameters for the other $S_4$ model\ec

\vss

As a result, in order to have enough parameters to fit the charged-lepton masses, $e_R$ must now be in {\bf 3}, not ${\bf 3'}$, nor a combination
of {\bf 1} and {\bf 2}. Other than that, things are pretty much the same as before except that $M_e$ is now no longer diagonal. To see the difference
between this case and TBM, let us illustrate again with the model in which $e_L, e_R, \nu_L, N_R$ all belong to {\bf 3}. In that case, using the
$F_0$-diagonal CG coefficient as before, but now a different $\bk{\f}$, we get
\be
M_e=\pmatrix{\ol c_1& \o\ol c_2+\ol c_3/\rt&\ol c_2-\o^2\ol c_3/\rt\cr   \ol c_2+\o^2\ol c_3/\rt & \ol c_1-\o\ol c_3/\rt&\o\ol c_2\cr
 \o\ol c_2-\ol c_3/\rt&\ol c_2&  \ol c_1+\o\ol c_3/\rt\cr},\labels{mec5}              \ee
where $\ol c_1=c_1/\rt,\ \ol c_2=c_2/\rs$, and $\ol c_5=c_5/\rd$. Since the $\bk{\chi}$'s are the same as in Table 4, the neutrino mass matrices
$M_\nu$ and $M_N$ are identical to those given in \eq{s4mn}, which are diagonalized by the TBM matrix $U_{0}$ of \eq{TBM}. If for convenience we
switch the second and the third columns, then it is $U_0'$ as explained below \eq{uc5}. This $U_0'$ however is not the mixing matrix because $M_e$
is not diagonal.

Since $M_e$ commutes with $F$, which is non-degenerate, the matrix $V$ in \eq{VV} that diagonalized $F$ must also diagonalize $M_e$. Indeed,
\be
V^\dagger M_e V=\pmatrix{\ol c_1+2\o^2\ol c_2&&\cr &\ol c_1-\o^2\ol c_2+\o\ol c_5&\cr &&\ol c_1-\o^2\ol c_2-\o\ol c_5\cr},\labels{mec5d}\ee
which gives rise to the charged-lepton mass-squard matrix $V^\dagger\ol M_e V=\diag(m_e^2, m_\mu^2, m_\tau^2)$, with
\be
m_e^2=(\ol c_1-\ol c_2)^2+3\ol c_2^2,\quad m^2_{\mu,\tau}=\ol c_1^2+\ol c_2^2+\ol c_5^2+\ol c_1\ol c_2\mp(\ol c_1-\ol c_2)\ol c_5.\ee
The neutrino mixing matrix is now given by $U=V^\dagger U_0'$, whose explicit result is already given in \eq{uc5} and \eq{uc51}.
\en

\subsection{$\bf A_4$}

We shall discuss for this group both the full-mixing and the partial-mixing models. For full-mixing, $\bk{\chi}$ 
is the simultaneous invariant eigenvector of $G_1, G_2$, and $G_3$, and is denoted by $\bk{\chi_{all}}$. 
For partial mixing defined by $G_i$, $\bk{\chi}$
 is the invariant eigenvector of that $G_i$, denoted as $\bk{\chi_{i}}$. Table 8 exhibits the IR of the residual operators and the vacuum alignments.

\vss
$$\ba{|c|c|c|c|c|}\hline
{\rm IR}&{\bf 1}&{\bf 1'}&{\bf 1''}&{\bf 3}\cr\hline
F&1&\o&\o^2&\diag(1,\o,\o^2)=F_0\cr\hline
\bk{\f}&1&0&0&(1,0,0)\cr\hline\hline
G_1&1&1&1&F_0G_{20}F_0^2:=G'_{10}\cr\hline
\bk{\chi_1}&1&1&1&(1,\o,\o^2)\cr\hline\hline
G_2&1&1&1&G_{20}\cr\hline
\bk{\chi_2}&1&1&1&(1,1,1)\cr\hline\hline
G_3&1&1&1&F_0^2G_{20}F_0:=G'_{30}\cr\hline
\bk{\chi_3}&1&1&1&(1,\o^2,\o)\cr\hline\hline
\bk{\chi_{all}}&1&1&1&(0,0,0)\cr
\hline\ea$$

\vss
\bc Table 8. Irreducible representations and invariant eigenvectors of  $A_4$\ec

\vss
The parameter count for these models is given in Table 9.
$$\ba{|c|c|c|c|c|c|}\hline
{\rm RHF}&&{\bf 1}&{\bf 1'}&{\bf 1''}&{\bf 3}\cr\hline
{\rm flavons}&&{\bf 3}&{\bf 3}&{\bf  3}&{\bf 1, 1', 1'', 3_S, 3_A}\cr\hline\hline
e\ {\rm param}&&1&1&1&3\cr\hline
{\rm typeII}\ \nu&G_1&-&-&-&4\cr
&G_2&-&-&-&4\cr
&G_3&-&-&-&4\cr
&all&-&-&-&3\cr\hline
{\rm typeI}\ \nu&G_1&1+1&1+1&1+1&5+4\cr
&G_2&1+1&1+1&1+1&5+4\cr
&G_3&1+1&1+1&1+1&5+4\cr
&all&0+0&0+0&0+0&3+3\cr
\hline\ea$$

\bc Table 9. Number of free parameters for the full-mixing and partial-mixing $A_4$ model\ec

\vss
We see once again that no matter how $e_R$ is assigned,  there are always just enough parameters to fit the charged-lepton masses. 
For type-II neutrinos, after using three to fit the neutrino masses, there is one left over to fit the remaining mixing. It is
this extra parameter that can turn a partial-mixing mode of $A_4$ into a TBM, thereby promoting the $A_4$ symmetry to a $S_4$ symmetry.
In the case of full mixing, listed under `all', the mixing matrix is completely determined to be the Cabibbo-Wolfenstein matrix, so
 the number of parameters is just enough to fit the neutrino masses. For type-I mixing, as before, there are more parameters than necessary
to fit the active neutrino masses, and the remaining mixing if in the partial-mixing mode, so the extra ones can be applied to tune the properties
in the heavy neutrino sector.

\section{Higgs Potential of $\bf A_4$}
We established in Secs.~II and III a connection between the mixing matrix and horizontal symmetry, purely on symmetry grounds,
without  the presence of Higgs nor a Lagrangian. 
We also saw in Sec.~IV that if 
flavons are introduced in a Lagrangian, then their vacuum alignments must be invariant eigenvectors of the residual
symmetry operators for the two formalisms to agree.
In a dynamical model, vacuum alignments  are obtained
by minimizing a Higgs potential, so for consistency we must find a suitable Higgs potential  to
yield just those alignments.

A completely generic Higgs potential contains so many $\G$-invariant terms that just about any vacuum alignment can
be obtained by adjusting the self-coupling parameters.
In order to confine the solutions to invariant eigenvectors, a rationale must be found
 to chop down the number of terms in the potential. The usual way to accomplish that is to impose an additional symmetry on top of $\G$.

Take $\G=A_4$ for example. 
It contains 
four irreducible representations, ${\bf 1, 1', 1'', 3}$.
Since the only non-trivial alignment occurs in {\bf 3}, let us consider
a renormalizable Higgs potential $V(\q,\q^*)$ of the complex triplet flavon field $\q=(a_1,a_2,a_3)^T$.
It turns out that if $V$ contains only $\q^*\q$ and $\q^*\q^*\q\q$ terms, without 
terms like $\q\q,\ \q^*\q\q, \q^*\q\q\q$, etc., 
then its minimization would
yield just the invariant eigenvectors of $A_4$. Such a potential can be arrived at by assigning $\q$
a non-trivial $U(1)$ quantum number, $\q^*$ the opposite quantum number, and imposing an additional
$U(1)$ symmetry on $V$. A suitable $Z(N)$ would do as well.

In addition to showing that the solution to its equations of motion are invariant eigenvectors, 
we shall also show that alignments for $A_4$ members of the same class have the same Higgs energy.
These general properties 
remain valid if $A_4$ is replaced by any simply reducible group $\G$, as we will see
in the next section.

Such a potential has been considered before \cite{A4pot}. For
the quartic terms, if we choose to couple the two $\q$'s first, and separately the two $\q^*$'s,
then we obtain four terms depending on whether $\q\q$ are coupled to ${\bf 1, 1', 1''}$, or {\bf 3}.
Using $\P^\a$ to denote the projection operator for irreducible representation $\a$, the most
general $V$ is 
\be V=
\bk{\q\q|f_1\P^1+f_{1'}\P^{1'}+f_{1''}\P^{1''}+f_3\P^3|\q\q}-\mu^2\bk{\q|\q},\labels{V}\ee
where $f_\a$ are the coupling constants.

With the help of explicit $A_4$ CG coefficients given in Appendix A,
$V$ can be written as 
\be V={1\over 3}\left(f_1|y_1|^2+f_{1'}|y_{1'}|^2+f_{1''}|y_{1''}|^2+2f_3(k^2-|y_3|^2)\right)-\mu^2k, \labels{V1}\ee
where
\be y_1&=&a_1^2+2a_2a_3,\quad y_{1'}=a_3^2+2a_1a_2,\quad y_{1''}=a_2^2+2a_1a_3,\nn\\
y_3&=&a_1a_2^*+a_2a_3^*+a_3a_1^*,\quad k=|a_1|^2+|a_2|^2+|a_3|^2. \labels{yk}\ee
The equations of motion are obtained by setting $\p V/\p a_i=0$ and $\p V/\p a_i^*=0$. In this way,
we obtain the following equation and its complex conjugate:
\be
0&=&f_1y_1\pmatrix{ a_1^*\cr a_3^*\cr a_2^*\cr}  +f_{1'}y_{1'}\pmatrix{ a_2^*\cr a_1^*\cr a^*_3\cr}
+f_{1''}y_{1''}\pmatrix{ a^*_3\cr a^*_2\cr a^*_1\cr} +
f_3\left[ 2k\pmatrix{ a_1\cr a_2\cr a_3\cr} 
-y_3\pmatrix{ a_2\cr a_3\cr a_1\cr} -y_3^*\pmatrix{  a_3\cr a_1\cr a_2\cr}\right]\nn\\ \nn\\ \nn\\
&-&
{3\over 2}\mu^2\pmatrix{ a_1\cr a_2\cr a_3\cr}. \labels{em}\ee
It is much easier to solve this cubic equation of $a_i$ by first converting it to a quadratic equation
in $y_a$ and $k$. To accomplish that, we multiply \eq{em} on the left respectively by
$(a_1^*, a_2^*, a_3^*),\ (a_3^*, a_1^*, a_2^*),\ (a_2^*, a_3^*, a_1^*)$ to get
\be
&&f_1|y_1|^2+f_{1'}|y_{1'}|^2+f_{1''}|y_{1''}|^2
+2f_3(k^2-|y_3|^2)-(3/2)\mu^2 k=0,\labels{qem1}\\
&&f_1y_1y_{1''}^*+f_{1'}y_{1'}y_{1}^*+f_{1''}y_{1''}y_{1'}^*
+f_3(ky_3^*-y_3^2)-(3/2)\mu^2 y_3^*=0, \labels{qem2}\\
&&f_1y_1y_{1'}^*+f_{1'}y_{1'}y_{1''}^*+f_{1''}y_{1''}y_{1}^*
+f_3(ky_3-y_3^{*2})-(3/2)\mu^2 y_3=0.\labels{qem3}\ee
Substituting \eq{qem1} into \eq{V1}, we can simplify the Higgs energy expression to
\be V=-{1\over 2}\mu^2k.\labels{V2}\ee
Subtracting the complex conjugate of \eq{qem3} from \eq{qem2}, and ask the equality
to be true for all $f_a$, we obtain
\be y_1y_{1''}^*=y_1^*y_{1'}=y_{1'}^*y_{1''}.\labels{yy}\ee
Let us now proceed to solve \eq{yy}.

\begin{enumerate}
\item 
 If $y_1=0$, then $y_{1'}$ or $y_{1''}$ must vanish. More generally, if one of the $y_s\ (s=1, 1', 1'')$
vanishes, then one or both of the other two must also vanish. In this way, we obtain four possible solutions:
$y:=[y_1, y_{1'}, y_{1''}]\propto [0,0,0], [1,0,0], [0,1,0], [0,0,1]$. 
\item If none of the $y_s$ vanish, then it follows from \eq{yy} that $|y_1|=|y_{1'}|=|y_{1''}|$. Dividing \eq{yy} by $y_1y_1^*$, 
we get $(y_{1''}/y_1)^*=(y_{1'}/y_1)=(y_{1''}/y_1)(y_{1'}/y_1)^*$. Hence $(y_{1''}/y_1)^3=1, (y_{1'}/y_1)^3=1$,
and $(y_{1''}/y_1)=(y_{1'}/y_1)^2$. The solutions are therefore $y=y_1[1,1,1],\ y_1[1,\o,\o^2]$, and $y_1[1,\o^2,\o]$.
\end{enumerate}

Substituting these solutions of $y_s$ into \eq{yk}, we can obtain the solutions of $\q=(a_1,a_2,a_3)$. Because of the $U(1)$
invariance, the solutions below can always be multiplied by an
arbitrary phase factor of unit modulus.
\begin{enumerate}
\item 
	\bn
	\i Suppose $y_{1'}=y_{1''}=0$. 
		\begin{enumerate}
		\item One obvious solution is $a_2=a_3=0$. In that case, $a_1=0$ if and
		only if $y_1=0$.  \vss

		The solution $\q=(0,0,0)$ gives rise to $k=y_3=0$, and a Higgs energy $V=0$ from \eq{V2}. Obviously it
		also satisfies the equation of motion \eq{em}.  \vss

		The solution $\q=\sqrt{k}(1,0,0)$ gives rise to $y_3=0$ and $y_1=k$. It satisfies \eq{em} and \eq{qem1} if 
		$k=3\mu^2/2(f_1+2f_3)$. The Higgs energy is then given by \eq{V2} to be $V=-3\mu^4/2(f_1+2f_3)$.

		\item There is also a solution with $a_2\not=0$. In that case, $y_{1'}=y_{1''}=0$ implies
		$a_1=-a_3^2/2a_2$ and $a_2^2=a_3^3/a_2$. The latter yields $a_3=a_2\o^m$ for $m=0,1$, or $2$, and the former
		demands $a_1=-a_2\o^{2m}/2$. From \eq{yk} we now obtain $y_1=9a_2^2\o^m/4,\ k=9|a_2|^2/4$, and $y_3=0$. We can now
		check that \eq{em} is satisfied provided $k=3\mu^2/2(f_1+2f_3)$, 
		so the Higgs energy $V=-3\mu^4/2(f_1+2f_3)$ is degenerate with
		that of the $\q=\sqrt{k}(1,0,0)$ solution. \vss
 
		This solution (actually three solutions corresponding to $m=0,1,2$) 
		can then be written as $\q=(\sqrt{k}/3)(1,-2\o^m,-2\o^{2m})$, or more explicitly, $\q=(\sqrt{k}/3)(1, -2, -2),
		(\sqrt{k}/3)(1,-2\o, -2\o^2), (\sqrt{k}/3)(1, -2\o^2, -2\o)$.
		\end{enumerate}
	\item  The solutions for $y_1=y_{1''}=0,\ y_{1'}\not=0$ can be similarly obtained. It leads to the solutions
	$\q=\sqrt{k}(0,1,0)$, $(\sqrt{k/3})(-2,1,-2)$, $(\sqrt{k}/3)(-2\o, 1, -2\o^2)$, and $(\sqrt{k}/3)(-2\o^2, 1, -2\o)$. The values
	for $k$ and $V$ are respectively $k=3\mu^2/2(f_{1'}+2f_3)$ and $V=-3\mu^4/2(f_{1'}+2f_3)$
	\item The solutions for $y_1=y_{1'}=0,\ y_{1''}\not=0$ are
	$\q=\sqrt{k}(0,0,1)$, $(\sqrt{k/3})(-2,-2, 1)$, $(\sqrt{k}/3)(-2\o, -2\o^2, 1)$, $(\sqrt{k}/3)(-2\o^2,  -2\o, 1)$. The values
	for $k$ and $V$ are respectively $k=3\mu^2/2(f_{1''}+2f_3)$ and $V=-3\mu^4/2(f_{1''}+2f_3)$.
	\i In short, the solutions of (b) and (c) are obtained from (a) with appropriate permutations.
	
	\end{enumerate}
\item 	\bn
	\i $y=y_1[1,1,1]$.  From \eq{yk}, we need to have 
	$a_1^2+2a_2a_3=a_3^2+2a_1a_2=a_2^2+2a_1a_3$.
	The first equality leads to $(a_1-a_3)(a_1+a_3-2a_2)=0$, and the second equality leads to 
	$(a_2-a_3)(a_2+a_3-2a_1)=0$. The solution is
	$a_1=a_2=a_3$, hence $\q=(\sqrt{k/3})(1,1,1)$. Using \eq{yk}, we get $y_1=y_{1'}=y_{1''}=y_3=k$.\vss

	We can check that \eq{em} is satisfied provided $k=3\mu^2/2(f_1+f_{1'}+f_{1''})$. The Higgs energy is then 
	$V=-3\mu^4/4(f_1+f_{1'}+f_{1''})$.

	\i $y=y_1[1,\o,\o^2]$.  From \eq{yk}, we need to have 
	$a_1^2+2a_2a_3=\o^2(a_3^2+2a_1a_2)=\o(a_2^2+2a_1a_3)$.
	The first equality leads to $(a_1-\o a_3)(a_1+\o a_3-2\o^2 a_2)=0$, and the second equality leads to 
	$(a_2-\o^2 a_3)(a_2+\o^2 a_3-2\o a_1)=0$. The solution is
	$a_1=\o^2 a_2=\o a_3$, hence $\q=(\sqrt{k/3})(1,\o,\o^2)$. Using \eq{yk}, we get $y_1=\o^2 y_{1'}=\o y_{1''}=y_3=k$.\vss

	We can check that \eq{em} is satisfied provided $k=3\mu^2/2(f_1+f_{1'}+f_{1''})$. The Higgs energy is then 
	$V=-3\mu^4/4(f_1+f_{1'}+f_{1''})$.

	\i $y=y_1[1, \o^2, \o]$. The solution is $\q=(\sqrt{k/3})(1,\o^2,\o)$, with $k=3\mu^2/2(f_1+f_{1'}+f_{1''})$. The Higgs energy is t
	$V=-3\mu^4/4(f_1+f_{1'}+f_{1''})$.
	\en

\item In summary, the solutions can be classified according to their energy into five categories:
	\bn
	\i $V=0$. In this case $\q=(0,0,0)$.
	\i $V=-3\mu^4/4(f_1+2f_3)$. In this case $\q=\sqrt{k}(1,0,0)$, $(\sqrt{k}/3)(1,-2,-2)$, $(\sqrt{k}/3)(1,-2\o,-2\o^2)$, or
	$(\sqrt{k}/3)(1,-2\o^2,-2\o)$, with $k=3\mu^2/2(f_1+f_3)$.
	\i $V=-3\mu^4/4(f_{1'}+2f_3)$. In this case $\q=\sqrt{k}(0,1,0)$, $(\sqrt{k}/3)(-2,1,-2)$, $(\sqrt{k}/3)(-2\o,1,-2\o^2)$, or
	$(\sqrt{k}/3)(-2\o^2,1,-2\o)$, with $k=3\mu^2/2(f_{1'}+f_3)$.
	\i $V=-3\mu^4/4(f_{1''}+2f_3)$. In this case $\q=\sqrt{k}(0,0,1)$, $(\sqrt{k}/3)(-2,-2,1)$, $(\sqrt{k}/3)(-2\o,-2\o^2,1)$, or
	$(\sqrt{k}/3)(-2\o^2,-2\o,1)$, with $k=3\mu^2/2(f_{1''}+f_3)$.
	\i $V=-3\mu^3/4(f_1+f_{1'}+f_{1''})$. In this case $\q=(\sqrt{k}/3)(1,1,1)$, $(\sqrt{k}/3)(1,\o,\o^2)$, or 
	$(\sqrt{k}/3)(1,\o^2,\o)$, with $k=3\mu^2/2(f_1+f_{1'}+f_{1''})$.
	\en
\end{enumerate}

In particular, if $f_1, f_{1'}, f_{1''}, f_3$ are all positive, energetics dictates that the solution $\q=(0,0,0)$ is disfavored. 
This is the solution that leads to the $A_4$ full-mixing matrix of Cabibbo and Wolfenstein, rather than the experimentally correct mixing
of the tribimaximal type, obtainable using the first solution of (b) in the charged lepton sector, and the first solution of
(e) in the neutrino sector.

Moreover, if $f_1< f_{1'}$ and $f_{1''}$, then it is (1,0,0) etc. that is energetically favored, and not (0,1,0), (0,0,1), etc.

In this way, we see that there is a rational why $A_4$ might prefer partial mixing, which may lead to a tribimaximal matrix, rather
than a full mixing that leads to a Cabibbo-Wolfenstein matrix. 

We proceed now to discuss the connection between these vacuum alignment solutions and the residual symmetry operators. Table 10 shows
the vacuum alignment solutions above, and the operator for which the the alignment is an invariant eigenvector. $F, G_1', G_2, G_3'$
are the generators of $A_4$  in Table 8, but with the subscript 0 omitted.

$$\begin{array}{|c|c|}\hline
alignment&operator\\ \hline
(1,0,0)&F,F^2\\
(1,-2,-2)&FG_2F,F^2G_2F^2\\
(1,-2\o,-2\o^2)&F^2G_2,G_2F\\
(1,-2\o^2,-2\o)&G_2F^2,FG_2\\ \hline
(1,1,1)&G_2\\
(1,\o,\o^2)&G_1'=F^2G_2F\\
(1,\o^2,\o)&G_3'=FG_2F^2\\ \hline
(0,0,0)&(G_1',G_2,G_3')\\
\hline\end{array}$$

\begin{center} Table 10. Dynamical alignments are invariant eigenvectors of appropriate operators\end{center}
\vskip.5cm
Here are some remarks concerning Table 10.
\bn
\i The first seven solutions in the left column are the invariant eigenvectors of the 11 operators in the right column,
each of them  responsible for a partial-mixing. 
Together with the identity operator, these twelve operators together form the $A_4$ group. 
\i The last solution is the simultaneous eigenvector of $G_1', G_2$, and $G_3'$ relevant for full-mixing.
\i Solutions under 3(c) and 3(d) are not listed in the table because they are obtained from 3(b) by permutations. 
They are the invariant eigenvectors of the same $A_4$ members expressed in a similarly permuted basis.
\i The first four solutions (3(b)) have the same energy, and the corresponding operators are all in class $\C_3$ or $\C_4$ of 
Table 2. The next three solutions (3(e)) also have the same energy, and the corresponding operators are all in class $\C_2$.
In other words, vacuum alignments for operators in the same class always have the same energy.
\en

\section{Group Theory and Dynamics}

We saw in the last section that, barring permutations that serve only to relabel the neutrino mass eigenstates,
every  minimizing solution of the scalar potential $V$ is an invariant eigenvector of some group element of $A_4$.
Moreover, conjugate group elements  give rise to identical energy.

 We shall present another derivation of these facts in the first subsection below. This alternative derivation
relies more on the general property of groups, and is therefore more amenable  to generalization.
The generalization will be carried out in the second subsection. 

The following simple fact will be used in this alternative approach.
\vss
\noindent{\bf Lemma}:\quad Let $g^\a$ be the $\a$ irreducible representation of $g\in\G$. Let $\f^\a$ be an eigenvector of $g^\a$ with
eigenvalue $\l^\a$. Let $\Phi^\rho:=(\f^{\a_1}\f^{\a_2}\cdots \f^{\a_n})^\r$ denote 
a state in the irreducible representation $\r$ obtained by coupling the $n$ states $\f^{\a_j}$. Then either $\Phi^\r=0$, or it
is an eigenvector of $g^\r$ with eigenvalue $\Lambda=\prod_{j=1}^n\l^{\a_j}$.
\vss
\noindent{\bf Proof}:\quad This follows immediately from the fact that $g^\r(\prod_{i}\f^{\a_i})^\r=(\prod_{i}(g^{\a_i}\f^{\a_i}))^\r$.
\vss\vss\vss
\noindent The following is an immediate corollary of Lemma 1.

\subsection{Another derivation of the  solution of ${\bf U(1)\x A_4}$}

An alternative derivation of the result of the last section is presented here. It has the advantage of being easily generalizable 
from $A_4$ to a general group $\G$.

The main idea  is the following. Start as before from the Higgs potential
$V=\sum_\a f_{\ol\a}\bk{\q\q|\P^{\ol\a}|\q\q}-\mu^2\bk{\q|\q}$. Its minimization is determined by six vectorial equations of motion,
$\p V/\p \q_c=0$ and $\p V/\p \q_c^*=0$. The last three is the complex conjugate of the first three so they need not be considered
separately. 
We will show, using the Lemma proven above, that if $\q$ is an invariant eigenvector of an element of $A_4$, then
every non-zero term in the three equations of motion is proportional to $\q$, with a scalar coefficient common to
the three equations. $\q$ is then a solution of the equations of motion if we set this scalar coefficient to be zero,
a condition that would be used to determined the normalization of the solution $\q$. Once $\q$ is known, its energy can be computed
from the above expression of $V$.

Here are the details. The idea is simple but unfortunately the notations are a bit cumbersome.  

$A_4$ has four irreducible representations, ${\bf 1, 1', 1'', 3}$. Other than the antisymmetric coupling ${ 3\times 3\to 3_A}$
which does not appear in \eq{V}, and will therefore be forgotten from now on,
each irreducible representation $\ol\a$ occurs at most once in the decomposition of $\b\x\g$.
The Clebsch-Gordan coefficients $\bk{\ol\a\ \ol a|\b b,\g c}$ for $\b\x\g\to\ol\a$ 
can be expressed in terms of the
$3j$-symbol,  $\bk{\ol\a\ \ol a|\b b,\g c}=\sqrt{[\a]}
{\footnotesize \pmatrix{\a&\b&\g\cr  a&b&c\cr}}$, where
$[\a]$ is the dimension of the irreducible representation $\a$, and the $3j$-symbol is completely symmetric in its columns.
The state $\ket{\ol\a\ \ol a}$ is the complex-conjugated state of $\ket{\a a}$. See Appendix A for further details.

Consider a typical quartic term in the real Higgs potential $V=\sum_\a f_{\ol\a}\bk{\q\q|\P^{\ol\a}|\q\q}-\mu^2\bk{\q|\q}$, 
\be
Q^\a:&=&\bk{\q\q|\P^{\ol\a}|\q\q}=[\a]\sum_{a,b,c,e,f}\bra{\q_e}\bra{\q_f}{\footnotesize\pmatrix{3&3&\a\cr e&f& a}^*}\ \
{\footnotesize\pmatrix{\a&3&3\cr  a&b&c\cr}}\ket{\q_b}\ket{\q_c}.\labels{palpha}\ee
Its contribution to the equation of motion is 
\be
{\p Q^\a\over\p \q_c}&:=&Q^\a_c=2[\a]\sum_{a,b,c,e,f}\bra{\q_e}\bra{\q_f}{\footnotesize\pmatrix{3&3&\a\cr e&f& a}^*}\ \
{\footnotesize\pmatrix{\a&3&3\cr  a&b&c\cr}}\ket{\q_b},\labels{dpalpha}\ee
and its complex conjugate $\p Q^\a/\p\q_c^*$.
In these formulas, $\ket{\q_b}$ stands for the $b$th component of a three-dimensional complex column vector $\q$,
and $\bra{\q_e}$ the complex conjugate of $\ket{\q_e}$. 

We will use $g^\r$ to denote IR$\r$ of a $g\in A_4$.  Suppose $g^{\a}$ has an orthonormal set of 
eigenvectors $\f^{\a m}\ (m=1,2,\cdots,[\a])$, with eigenvalues $\l_m$,
then $\d_{ad}=\sum_m\f^{\a m *}_a\f^{\a m}_d$, so \eq{palpha} and \eq{dpalpha} can
be rewritten as
\be
Q^\a&=&[\a]\sum_m Y^{\a m *}Y^{\a m},\labels{beta}\\
Q^\a_c&=&2[\a]\sum_mY^{\a m *}\Phi^{\a m}_c:=2[\a]\sum_m Q^{\a m}_c,\labels{dbeta}\ee
where
\be
Y^{\a m}&=&\sum_{abc}{\footnotesize\pmatrix{\a&3&3\cr  a&b&c\cr}}\phi^{\a m}_{ a}\q_b\q_c,\labels{Y}\\
\Phi^{\a m}_c&=&\sum_{ab}{\footnotesize\pmatrix{\a&3&3\cr  a&b&c\cr}}\phi^{\a m}_{ a}\q_b.\labels{dY}
\ee
It follows from the symmetry of the $3j$-symbol that $Y^{\a m}$ transforms as {\bf 1}, and $\Phi^{\a m}$ transforms as {\bf 3}.

We can now start working towards the goal of showing that invariant eigenvectors of $g$ are solutions to the equations of motion derived from $V$.

Let $\q$ be an invariant eigenvector of $g$,
{\it i.e.}, $g^3\q=\q$. Since $g^{ \a}\f^{\a m}=\l_m\f^{\a m}$, according to the Lemma, either 
$Y^{\a m}=0$, or $g^1Y^{\a m}=\l_m Y^{\a m}$. However, since $g^1=1$ for every $g$ in the identity representation {\bf 1}, 
we must have
$\l_m=1$ in order for $Y^{\a m}\not=0$. 

The Lemma also implies that either $\Phi^{\a m}=0$,  or  $g^3\Phi^{\a m}=\l_m\Phi^{\a m}$. Consequently,
if $Q^{\alpha m}_c\not=0$, meaning both $Y^{\a m}$ and $\Phi^{\a m}$ non-zero, then it is necessary for $\Phi^{\a m}$ to be an invariant eigenvector of $g^3$. 
We know from Table 8 that every $g^3$ has one and only one invariant eigenvector, hence $\Phi^{\a m}=\kappa^\a \q$ for some $\kappa^\a\not=0$, 
if $Q^{\a m}_c\not=0$.

The importance of this observation is the following. There are six equations of motion, namely, the following three ($c=1,2,3$), and their
three complex conjugates:
\be \sum_{\a=11'1''3}2[\a]f_\a \sum_m Q^{\a m}_c-\mu^2\q_c=\sum_{\a=11'1''3}2[\a]f_\a \sum_m Y^{\a m *}\Phi^{\a m}_c-\mu^2\q_c=0.\labels{emq}\ee
From the observation above, we see that if $\q$ is an invariant eigenvector of $g^3$, then every non-zero term in \eq{emq} is proportional
to $\q$, hence {\it all} the six equations of motion are satisfied provided the single scalar equation
\be \sum_{\a=11'1''3}2[\a]f_\a  Y^{\a m *}\kappa^\a-\mu^2=0\labels{norm}\ee
is obeyed. Note that the single $m$ in the equation is that one for which $\Phi^{\a m}$ is the invariant eigenvector of $g^3$.
Since every quantity in this equation is known, except the overall normalization of $\q$, this then is an equation to determine the
normalization of $\q$. For that reason we shall sometimes refer to \eq{norm} as the {\it normalization equation}.

Knowing that $\q$ is an invariant eigenvector of $g$ with known normalization computed in \eq{norm}, we can substitute this solution
into the expression of $V$ to compute its energy.

To summarize, we have shown that the invariant eigenvector $\q$ of every group element $g$ of $A_4$ is a dynamical vacuum alignment,
satisfying the equation of motion derived from the Higgs potential $V$. We will now show that conjugate elements in $A_4$ give rise
to the same energy $V$.

Let $g'=hgh^{-1}$ be an element conjugate to $g$. 
If $g^3\q=\q$, then $g^{'3}\q'=\q'$ for $\q'=h^3\q$. Similarly, $g^{'\a}\f^{'\a m}=\l_m\f^{'\a m}$
if $g^\a\f^{\a m}=\l_m\f^{\a m}$ and $\f^{'\a m}=h^\a\f^{\a m}$. If we denote the quantities in \eq{Y} and \eq{dY} built up of the
primed quantities by $Y^{'\a m}$ and $\Phi^{'\a m}_c$ respectively, then $Y^{'\a m}=Y^{\a m}$ and $\Phi^{'\a m}=h^3\Phi^{\a m}$. 
Furthermore, if $\Phi^{\a m}=\kappa^\a\q$, then $\Phi^{'\a m}=\kappa^\a\q'$, with the same $\kappa^\a$, 
For that reason, the normalization equation \eq{norm} is identical for the primed as the unprimed quantities,
so $\q$ and $\q'$ have the same normalization, thus both solutions give rise to the same energy $V$.

The eigenvalues of $G^3_i$ are $+1, -1, -1$, and the eigenvalues of all other $g^\r$ can be obtained from Table 8. Using
this information, we can make a closer comparison between the present derivation and the explicit calculation in Sec.~V.

If $g$ is an order-3
element, then $g^\a\not=1$ in the $\a={\bf 1', 1''}$ representations, hence $Y^{\a m}=0$ for $\a={\bf 1', 1''}$. This corresponds to
the $y_{1'}=y_{1''}=0$ solutions in the previous section. Namely, the first four rows of Table 10, or solution 3(b) before that table.

If $g$ is order-2, for the singlet representations $\a={\bf 1, 1', 1''}$, every $g^\a=+1$, hence according to the Lemma, none of these
three $Y^{\a m}$ (there is only one $m$) have to vanish, and we can take $\f^{\a m}=1$ in all cases. This corresponds
 to the solutions with $|y_1|=|y_{1'}|=|y_{1''}|\not=0$
in the last section, given by 3(e) and the next three rows of Table 10.

\subsection{Generalization to other groups $\G$}

The alternative derivation
 for $A_4$ discussed above is so general that it can be easily adapted to many other groups $\G$. For technical simplicity we
shall confine ourselves here to the simply reducible groups. The validity of the theorem below certainly goes beyond the simply reducible
groups, as $A_4$ itself is not simply reducible. For those groups, the same idea should
still be applicable, though generalization and some modification in details may be necessary.

A {\it simply reducible group} is a group in which (i) $g$ and $g^{-1}$ are always in the same class, and (ii) in the decomposition of the Kronecker product $\b\x\g$
of any two irreducible representations $\b$ and $\g$, every irreducible representation $\ol\a$ occurs at most once. A consequence
of (i) is that all characters have to be real. $A_4$ is not
simply reducible because it contains complex characters, but groups like $SU(2), SO(3)$, $S_3$, $S_4$, and the quaternion group,
 are simply reducible.

The main property we need from a simply reducible group is the existence of a symmetric or antisymmetric $3j$-symbol. As before, if
$\bk{\ol\a\ \ol a|\b b,\g c}$ is the Clebsch-Gordan coefficient, then the $3j$-symbol is defined by
\be \bk{\ol\a\ \ol a|\b b,\g c}=\sqrt{[\a]}{\footnotesize\pmatrix{\a&\b&\g\cr  a&b&c\cr}}\labels{3j}\ee
is either symmetric or antisymmetric upon the interchange of columns. Here
$(\ol\a\ \ol a)$ is the complex-conjugate representation of $(\a a)$; whether the $3j$-symbol is symmetric or antisymmetric is a detail that does
not concern us.

Most of the arguments used in $A_4$ can now be carried through to prove

\noindent{\bf Theorem:}\quad Let $\q$ be a complex variable vector in the $\r$ irreducible representation of a simply reducible group $\G$,
carrying a non-trivial $U(1)$ quantum number opposite to that of $\q^*$. Writing the most general $(U(1)\x\G)$-invariant 
renormalizable real Higgs potential as 
\be
V=\sum_\a f_{\ol\a}\bk{\q\q|\P^{\ol\a}|\q\q}-\mu^2\bk{\q|\q},\labels{VG}\ee
in which the sum is carried over all irreducible representations, with $\P^\a$ being the projection operator and $f_\a$ the Yukawa
coupling constant of the $\a$ representation. Then every {\it unique} invariant eigenvector $\q$ of $g^\r$ for every $g\in\G$
is a solution of the equation of motion $\p V/\p\q_c=\p V/\p\q^*_c=0$. Moreover, 
 $g$'s in the same conjugacy class have the same
Higgs energy $V$.
\vskip.5cm
\noindent{\bf Proof:}\quad The proof is essentially a copy of the procedure discussed in the last subsection. The only thing 
we are not sure without knowing more details about $\G$ is whether every $g^\r$ has one and only one eigenvector with eigenvalue
$+1$. This is why the theorem is limited to those $g^\r$ with {\it unique} invariant eigenvectors. The only other difference is that
the $3j$-symbol may be antisymmetric under a column exchange rather than symmetric. The only place this symmetry is used
was in \eq{Y} and \eq{dY} to show that they transform as {\bf 1} and {\bf 3} respectively. Clearly an additional minus sign under
column exchange will not affect this property. Finally, whether $\q$ belongs to {\bf 3} or some other
multi-dimensional irreducible representation $\r$ really does not matter in the proof. 

With these preliminary remarks, let us proceed with the details.

Consider a typical quartic term in the real Higgs potential $V$, 
\be
Q^\a:&=&\bk{\q\q|\P^{\ol\a}|\q\q}=[\a]\sum_{a,b,c,e,f}\bra{\q_e}\bra{\q_f}{\footnotesize\pmatrix{\r&\r&\a\cr e&f& a}^*}\ \
{\footnotesize\pmatrix{\a&\r&\r\cr  a&b&c\cr}}\ket{\q_b}\ket{\q_c}.\labels{palphag}\ee
Its contribution to the equation of motion is given by
\be
{\p Q^\a\over\p \q_c}&:=&Q^\a_c=2[\a]\sum_{a,b,c,e,f}\bra{\q_e}\bra{\q_f}{\footnotesize\pmatrix{\r&\r&\a\cr e&f& a}^*}\ \
{\footnotesize\pmatrix{\a&\r&\r\cr  a&b&c\cr}}\ket{\q_b},\labels{dpalphag}\ee
together with its complex conjugate $\p Q^\a/\p\q_c^*$.

Suppose $g^{\a}$ has an orthonormal set of 
eigenvectors $\f^{\a m}\ (m=1,2,\cdots,[\a])$, with eigenvalues $\l_m$.
Then $\d_{ad}=\sum_m\f^{\a m *}_a\f^{\a m}_d$, so \eq{palphag} and \eq{dpalphag} can
be rewritten as
\be
Q^\a&=&[\a]\sum_m Y^{\a m *}Y^{\a m},\labels{betag}\\
Q^\a_c&=&2[\a]\sum_mY^{\a m *}\Phi^{\a m}_c:=2[\a]\sum_m Q^{\a m}_c,\labels{dbetag}\ee
where
\be
Y^{\a m}&=&\sum_{abc}{\footnotesize\pmatrix{\a&\r&\r\cr  a&b&c\cr}}\phi^{\a m}_{ a}\q_b\q_c,\labels{Yg}\\
\Phi^{\a m}_c&=&\sum_{ab}{\footnotesize\pmatrix{\a&\r&\r\cr  a&b&c\cr}}\phi^{\a m}_{ a}\q_b.\labels{dYg}
\ee
It follows from the symmetry of the $3j$-symbol that $Y^{\a m}$ transforms as {\bf 1}, and $\Phi^{\a m}$ transforms as ${\bf \r}$.

Suppose $\q$ is an invariant eigenvector of $g$,
{\it i.e.}, $g^\r\q=\q$, and suppose this is the only eigenvector of $g^\r$ with eigenvalue $+1$. 
Since $g^{ \a}\f^{\a m}=\l_m\f^{\a m}$, according to the Lemma, either 
$Y^{\a m}=0$, or $g^1Y^{\a m}=\l_m Y^{\a m}$. However, since $g^1=1$ for every $g$ in the identity representation {\bf 1}, 
we must have
$\l_m=1$ in order for $Y^{\a m}\not=0$. 

The Lemma also implies that either $\Phi^{\a m}=0$,  or  $g^\rho\Phi^{\a m}=\l_m\Phi^{\a m}$. Consequently,
if $Q^{\alpha m}_c\not=0$, it is necessary for $\Phi^{\a m}$ to be an invariant eigenvector of $g^\rho$. 
Since this is assumed to be unique, it follows that $\Phi^{\a m}=\kappa^\a \q$ for some $\kappa^\a$ 
if $Q^{\a m}_c\not=0$.

Let us look at the equation of motion:
\be \sum_{\a}2[\a]f_\a \sum_m Q^{\a m}_c-\mu^2\q_c=\sum_{\a}2[\a]f_\a \sum_m Y^{\a m *}\Phi^{\a m}_c-\mu^2\q_c=0.\labels{gemq}\ee
From the observation above, we see that if $\q$ is an invariant eigenvector of $g^\r$, then every non-zero term in \eq{gemq} is proportional
to $\q$, hence equation of motion are satisfied provided the single scalar equation
\be \sum_{\a}2[\a]f_\a  Y^{\a m *}\kappa^\a-\mu^2=0\labels{gnorm}\ee
is obeyed. Note that the single $m$ in the equation is that one for which $\Phi^{\a m}$ is the invariant eigenvector of $g^\r$.
Since every quantity in this equation is known, except the overall normalization of $\q$, this then is an equation to determine the
normalization of $\q$. For that reason we shall sometimes refer to \eq{gnorm} as the {\it normalization equation}.

This proves the first part of the theorem. What remains is to show that conjugate elements in $\G$ give rise
to the same energy $V$.

Let $g'=hgh^{-1}$ be an element conjugate to $g$. 
If $g^\rho\q=\q$, then $g^{'\r}\q'=\q'$ for $\q'=h^\r\q$. Similarly, $g^{'\a}\f^{'\a m}=\l_m\f^{'\a m}$
if $g^\a\f^{\a m}=\l_m\f^{\a m}$ and $\f^{'\a m}=h^\a\f^{\a m}$. If we denote the quantities in \eq{Yg} and \eq{dYg} built up of the
primed quantities by $Y^{'\a m}$ and $\Phi^{'\a m}_c$ respectively, then $Y^{'\a m}=Y^{\a m}$ and $\Phi^{'\a m}=h^\r\Phi^{\a m}$. 
Furthermore, if $\Phi^{\a m}=\kappa^\a\q$, then $\Phi^{'\a m}=\kappa^\a\q'$, with the same $\kappa^\a$, 
For that reason, the normalization equation \eq{norm} is identical for the primed as the unprimed quantities,
so $\q$ and $\q'$ have the same normalization, thus both solutions give rise to the same energy $V$.

\section{Conclusion}
We have discussed in some detail the group-theoretical
 connection between a neutrino mixing matrix and the horizontal symmetry of left-handed leptons.
Horizontal symmetry groups can be derived from mixing matrices, and mixing matrices can be obtained from horizontal groups, all without
invoking the Higgs or other dynamical mechanisms, as long as we assume some trace of the original symmetry remains to be found
in the mass matrices even after symmetry breaking. We have also shown that this formulation is completely consistent with the dynamical
approach, in that the mixing matrices derived from a symmetry group $\G$ can be obtained from a 
generic Higgs potential invariant under $U(1)\x\G$,
for many groups $\G$. Such a potential gives rise to vacuum alignments that are invariant eigenvectors of the group elements. 
An alignment should be assigned to the charged-lepton sector if the corresponding group element has order $\ge 3$, but to the
neutrino sector if the corresponding group element has order 2.
Within this rule, alignments can be further discriminated by their energies.
 
Other than Sec.~IV, in which some $S_4$ and $A_4$ dynamical models are constructed to illustrate the consistency of the two approaches, we
have not studied in detail in this article how physical models should be constructed. 
The models constructed in Sec.~IV may not be optimal, in that
mass hierarchy is not natural in those models, and  the successful
Koide mass relation of the charged-leptons \cite{KOIDE} is not explained. 
 More complicated dynamics including the presence of quadratic valon fields may be necessary. Furthermore,
although the $U(1)\x\G$ potential can give rise to all the desired vacuum alignments, and although their assignments
to the two sectors can be discriminated by
the order of the associated residual symmetry operator, we have not found a purely dynamical way to do so without invoking
the residual symmetry operators explicitly. That may call for the introduction
of driver valons \cite{ALTARELLI} or other dynamical mechanisms.
\newpage
\appendix
\section{}
In this appendix, we list the Clebsch-Gordon (CG) series of $A_4$ and its CG coefficients computed
in the representation of Table 8.

The CG series $\b\x\g\to\ol\a$ can be computed from Table 2 to be
$$\ba{|c||c|c|c|c|}\hline
&1&1'&1''&3\\ \hline\hline
1&1&1'&1''&3\\ \hline
1'&1'&1''&1&3\\ \hline
1''&1''&1&1'&3\\ \hline
3&3&3&3&1,1',1'',3_S,3_A\\ 
\hline\ea$$
\bc Table 11. Clebsch-Gordan series of $A_4$\ec
where the first row refers to IR$\g$, the first column refers to IR$\b$, and the table entries refer to
 IR$\ol\a$. The CG coefficients $\bk{\ol\a\ \ol a|\b b,
\g c}$ are zero if $(\ol\a,\b,\g)$ is not contained in an entry of the table. It is also zero if the product
of $F$-eigenvalues of $\ket{\b b}$ and $\ket{\g c}$ is not equal to that of $\ket{\ol\a\ \ol a}$. This 
last condition is a consequence of eq.~\eq{CGI} applied to $g=F$. If we use $A, B, C$ to denote the three states in $1, 1', 1''$,
and $D, E, G$ to denote the three states in $3$, then the $F$-eigenvalue of $A$ and $D$ is 1, that of $B$ and $E$ is $\o$, 
and that of $C$ and $G$ is $\o^2$. The eigenvalue rule is not shown in Table 11, but that is incorporated in the expanded 
Table 12.

$$\ba{|c||c|c|c|c|c|c|}\hline
&A&B&C&D&E&G\\ \hline\hline
A&A&B&C&D&E&G\\ \hline
B&B&C&A&E&G&D\\ \hline
C&C&A&B&G&D&E\\ \hline
D&D&E&G&A,D_S&B,E_S,E_A&C,G_S,G_A\\ \hline
E&E&G&D&B,E_S,E_A&C,G_S&A,D_S,D_A\\ \hline
G&G&D&E&C,G_S,G_A&A,D_S,D_A&B,E_S\\ \hline
\ea$$
\bc Table 12. Expanded Clebsch-Gordan series of $A_4$\ec

The complex conjugate $\ket{\ol\a\ \ol a}$ of a state $\ket{\a a}$ is defined to be the state with the complex-conjugated $F$-eigenvalue.
Thus $\ol X=X$ if $X=A, D$, but $\ol B=C,\ \ol C=B, \ol E=G, \ol G=E$.

We will also write the CG coefficient simply as $\bk{\ol X|Y,Z}$ to represent the more complicated $\bk{\ol\a\ \ol a|\b b,\g c}$.

The CG coefficients are related to  the $3j$-symbols 
via the equation
\be \bk{\ol\a\ \ol a|\b b,\g c}&=&\sqrt{[\a]}{\footnotesize\pmatrix{\a&\b&\g\cr a&b&c\cr}},\nn\\
\bk{\ol X|Y,Z}&=&\sqrt{[X]}\pmatrix{X&Y&Z\cr},
\labels{3j}\ee
where $[\a]$ is the dimension of IR$\a$, and $[X]$ is the dimension of the IR in which state $X$ belongs to. The $3j$-symbol
couples three states to a singlet; it vanishes unless the
product of the three $F$-eigenvalues is 1.

There are two $3j$-symbols when $\a=\b=\g=3$. The one  corresponding to $\ol\a= 3_A$ will be marked with a subscript
$A$, and the one corresponding to $\ol\a=3_S$ will carry no subscript.

$3j$-symbols are more convenient because of their symmetry properties.
The $3j$-symbol $(X\ Y\ Z)_A$ is antisymmetric upon the interchange of two columns, and all the other
$3j$-symbols are symmetric upon such an exchange.

The non-zero $3j$-symbol $(X\ Y\ Z)$ is equal to 1 when $X, Y, Z$ all belong to 1-dimensional IRs, is equal to $1/\rt$
if two of them belong to the 3-dimensional IR and the third one belongs to a 1-dimensional IR. Finally, when all three of them
belong to the 3-dimensional IR, then their values are
\be
(D\ E\ G)_A&=&-{1\over\rd}, \qquad (D\ E\ G)=-{1\over\sqrt{6}},\nn\\
 (D\ D\ D)&=&(E\ E\ E)=(G\ G\ G)=\sqrt{2\over 3}.\labels{a43j}\ee
It can be checked that the CG coefficients obey eq.~\eq{CGI} for $g=G_2$ as well.

\end{document}